\documentclass[aps,prd,twocolumn,amsmath,amssymb,showpacs,floatfix,superscriptaddress,nofootinbib]{revtex4-1}

\usepackage{natbib}
\usepackage{amsmath}
\usepackage{amssymb}
\usepackage{latexsym}
\usepackage{bm}   
\usepackage{graphicx,epstopdf}
\usepackage{epstopdf}
\DeclareGraphicsExtensions{.ps}
\usepackage{dcolumn} 
\usepackage{multirow}
\usepackage{appendix}
\usepackage{footnote}
\usepackage{tabularx,ragged2e,booktabs}

\newcolumntype{C}[1]{>{\Centering}m{#1}}

\usepackage{color}

\DeclareMathAlphabet\mathbfcal{OMS}{cmsy}{b}{n}
\definecolor{darkgreen}{cmyk}{0.85,0.2,1.00,0.35} 
\definecolor{purple}{cmyk}{0.5,1.0,0,0} 
\definecolor{darkblue}{cmyk}{1.0,1.0,0,0}
\usepackage{hyperref}

\hypersetup{
   colorlinks = true,
   allcolors = darkblue}
   
\newcommand{\before}{{\rm before}}	
\newcommand{\by}{{\rm by}}

\newcommand{\eff}{{\rm eff}}
\newcommand{\CIP}{{\rm CIP}}
\newcommand{\CMBavg}{{\rm CMB}}
\newcommand{\fsky}{f_{\rm sky}}

\def\beq{\begin{equation}}
\def\eeq{\end{equation}}
\def\bea{\begin{eqnarray}}
\def\eea{\end{eqnarray}}
\def\lsim{\mathrel{\raise.3ex\hbox{$<$\kern-.75em\lower1ex\hbox{$\sim$}}}}
\def\gsim{\mathrel{\raise.3ex\hbox{$>$\kern-.75em\lower1ex\hbox{$\sim$}}}}
\def\wigner#1#2#3#4#5#6{ \left( \begin{array}{ccc} #1 & #3 & #5
\\ #2 & #4 & #6 \\ \end{array} \right)}

\newcommand{\DD}{_L^{\Delta\Delta}}

\newcommand{\rsig}{r_D}

\newcommand{\bx}{{\bm x}}
\newcommand{\bl}{{\bm l}}
\newcommand{\bk}{{\bm k}}
\newcommand{\bK}{{\bm K}}
\newcommand{\bL}{{\bm L}}
\newcommand{\bn}{\hat{\bm n}}

\newcommand{\apjl}{Astrophys. J. Lett.}

\newcommand{\jcap}{JCAP.}

\begin{document}
	

\title{Compensated isocurvature perturbations in the curvaton model}

\author{Chen He}\email{chenhe@uchicago.edu}
\affiliation{Department of Physics, University of Chicago, Chicago, Illinois 60637, USA}

\author{Daniel Grin}
\affiliation{Kavli Institute for Cosmological Physics, Chicago, Illinois 60637, USA}
\affiliation{Department of Astronomy and Astrophysics, University of Chicago, Chicago, Illinois 60637, USA} 

\author{Wayne Hu}
\affiliation{Kavli Institute for Cosmological Physics, Chicago, Illinois 60637, USA}
\affiliation{Department of Astronomy and Astrophysics, University of Chicago, Chicago, Illinois 60637, USA} 
\affiliation{Enrico Fermi Institute, University of Chicago, Chicago, Illinois 60637, USA} 


\begin{abstract}
Primordial fluctuations in the relative number densities of particles, or isocurvature perturbations, are generally well constrained by cosmic microwave background (CMB) data. A less probed mode is the compensated isocurvature perturbation (CIP), a fluctuation in the relative number densities of cold dark matter and baryons. In the curvaton model, a subdominant field during inflation later sets the primordial curvature fluctuation $\zeta$. In some curvaton-decay scenarios, the baryon and cold dark matter isocurvature fluctuations nearly cancel, leaving a large CIP correlated with $\zeta$. This correlation can be used to probe these CIPs more sensitively than the uncorrelated CIPs considered in past work, essentially by measuring the squeezed bispectrum of the CMB for triangles whose shortest side is limited by the sound horizon. Here, the sensitivity of existing and future CMB experiments to correlated CIPs is assessed, with an eye towards testing specific curvaton-decay scenarios.  The planned CMB Stage 4 experiment could detect the largest CIPs attainable in curvaton scenarios with more than 3$\sigma$ significance.  The significance could improve if small-scale CMB polarization foregrounds can be effectively subtracted.  As a result, future CMB observations could discriminate between some curvaton-decay scenarios in which baryon number and dark matter are produced during different epochs relative to curvaton decay.  Independent of the specific motivation for the origin of a correlated CIP perturbation, cross-correlation of CIP reconstructions with the primary CMB can improve the signal-to-noise ratio of a CIP detection.  For fully correlated CIPs the improvement is a factor of $\sim$$2-$3.
\end{abstract}


\maketitle

\section{Introduction}
The measured cosmic microwave background (CMB) anisotropy power spectra~\cite{Hinshaw:2012aka, Ade:2015lrj} are consistent with adiabatic primordial fluctuations, initial conditions for which the relative {particle number densities} are spatially constant. Adiabatic perturbations arise in the simplest inflationary models,
where a single field drives inflation and sets the amplitude of perturbations in all species. If fluctuations in particle densities or quantum numbers observed today are actually set by fluctuations in more than one field, some fraction of the primordial fluctuations may be isocurvature (also known as entropy) perturbations, for which there are initial fluctuations in the relative {particle number densities}.

The second field may be an axion~\cite{Linde:1984ti} (and thus a dark-matter candidate), a curvaton~\cite{Linde:1984ti, Linde:1996gt, Langlois:2000ar,Lyth:2002my} (a field that is energetically subdominant during inflation but later sets the density fluctuations in standard-model species), or alternatively, inflation itself may be driven by multiple fields with different couplings to standard-model particles~\cite{Linde:1984ti}. Fluctuations very similar to isocurvature fluctuations may also arise in topological-defect models~\cite{Axenides:1983hj,Brandenberger:1993by,Langlois:2000ar,Lalak:2007vi}. 

Isocurvature fluctuations are defined by the entropy fluctuation
\begin{equation}
S_{i\gamma}=\frac{\delta n_{i}}{n_{i}}-\frac{\delta n_{\gamma}}{n_{\gamma}}\end{equation} between a species $i$ and the photons ($\gamma$); where $n_{i}$ denotes the background number density of the species; $\delta n_{i}$ its spatial fluctuation; and $i\in \left\{b,c,\nu,\gamma\right\}$, where $b$ denotes baryons, $c$ denotes cold dark matter (CDM), and $\nu$ denotes neutrinos. 
These isocurvature modes individually leave an imprint 
on the temperature and polarization power spectra of the CMB~\cite{Langlois:2000ar} and are highly 
constrained by current data~\cite{Ade:2015lrj}.

There is one joint combination of isocurvature fluctuations that largely escapes constraints.
If
\begin{equation}
S_{c\gamma} = -\frac{\rho_b}{\rho_c} S_{b\gamma}, \quad S_{\nu\gamma}=0,
\end{equation}
then the density perturbations carried by the two isocurvature modes cancel in this combination when both the
baryons and CDM are nonrelativistic. This is called a 
 compensated isocurvature perturbation (CIP).  
 
 At linear order, CIPs only affect observables through the difference in the baryon and CDM pressure and hence on scales comparable to the baryonic Jeans length \cite{Gordon:2002gv,Barkana:2005xu,2007PhRvD..76h3005L,Gordon:2009wx,2011JCAP...10..028K}.  For the CMB, these scales are deep into the damping tail and the regime of secondary anisotropy dominance, as well as beyond the beam scale of any foreseeable CMB experiment. CIPs thus do not induce an observable effect on the CMB at linear order \cite{2007PhRvD..76h3005L,Lewis:2002nc,Gordon:2002gv}. 
 
There are potentially observable signatures on the $21$ cm signature of neutral hydrogen at very high redshifts (in absorption).  Sufficiently sensitive measurements for a CIP detection, however, will require a futuristic space-based $21$ cm experiment with a baseline that dwarfs that of ongoing/upcoming 21 cm efforts like MWA/LOFAR/PAPER/SKA by an order of magnitude~\cite{Gordon:2009wx}.
 
On the other hand, since CIPs modulate the photon-baryon  and baryon-CDM ratios, they do impact the
CMB at higher order.	By modulating these quantities in space, 
CIPs change the two-point correlations between CMB multipole moments in a way that
allows their reconstruction \cite{Grin:2011tf}.   This fact was applied to the 
WMAP 9-year data in Ref.~\cite{Grin:2013uya} to set upper limits on the CIP amplitude
independently of their origin.
Similar limits follow from measurements of the gas fraction in massive galaxy clusters \cite{Holder:2009gd}. In these prior works, the CIP was not assumed to be correlated with the dominant adiabatic fluctuation. Here we consider an early-universe mechanism that generates CIPs correlated with
adiabatic fluctuations, yielding another detectable signature.

Correlated isocurvature fluctuations arise naturally in the curvaton model, in which the curvaton, a subdominant field during inflation later seeds the observed primordial curvature fluctuations~\cite{Lyth:2002my,Gupta:2003jc,Gordon:2002gv,2009JCAP...11..003E}. As different species and quantum numbers may be generated by, before, or after curvaton decay, there are mismatches in their number densities which lead to isocurvature fluctuations including correlated CIPs. 

We assess the sensitivity of CMB anisotropy measurements to correlated CIPs generated in various curvaton-decay scenarios and find that a CMB Stage 4 \cite{Abazajian:2013oma} experiment could yield a detection of the largest such CIPs with more than 3$\sigma$ significance. The significance could improve to $11\sigma$
if polarized foregrounds and systematics can be modeled sufficiently to make a cosmic-variance
limited measurement out to multipoles of $l=4000$. More generally, we find that cross-correlation of CIP reconstructions with the primary CMB can improve the signal-to-noise ratio for detection of fully correlated CIPs by a factor of $\sim$2$-$3 depending on the specific experiment. 
		
		We establish that our reconstruction method \cite{Grin:2011tf} relies on a separate-universe (SU) approximation, limiting its use to angular scales $L\lsim100$. This has little impact for the signal-to-noise ratio of CIP searches using completed CMB experiments, but ultimately limits CIP reconstruction from nearly cosmic-variance limited future CMB polarization experiments. We correct numerical errors in the reconstruction noise curves of Ref. \cite{Grin:2011tf}; these errors are ultimately negligible on the scales where the SU approximation is valid. We also update CIP estimators to include sample variance from CMB $B$-mode polarization, as well as covariance between CIP estimators based on off-diagonal correlations between different pairs of observables (e.g.\ temperature, $E$-mode polarization, and $B$-mode polarization).
				 				
		We begin in Sec.~\ref{sec:curvaton} by reviewing the predictions for the amplitude of isocurvature perturbations and their correlations with the adiabatic mode in nine curvaton-decay scenarios.  In Sec.~\ref{sec:reconstruction}, we examine the tools for CIP reconstruction and compute updated reconstruction noise spectra based on the methods from Ref.~\cite{Grin:2011tf}. 
	In Sec.~\ref{sec:sigma} we determine the sensitivity of future CMB experiments to curvaton-inspired correlated CIPs. We assess the improvements in signal-to-noise ratio made possible by cross-correlating the CIP reconstruction with CMB temperature and polarization maps. We conclude in Sec.~\ref{sec:conclude}. In Appendix~\ref{sec:wprop}, we show that our reconstruction methods are limited to CIP modes that are larger than the sound horizon at recombination. In Appendix~\ref{sec:correct_noise} we discuss differences with the reconstruction results of Ref.~\cite{Grin:2011tf}.

	\section{Correlated CIPs in Curvaton Models}
	
	\label{sec:curvaton}
	
\subsection{General considerations}

The curvaton $\sigma$ is a light spectator scalar field during inflation and starts to oscillate when the Hubble scale $H$ approaches the curvaton mass $m_{\sigma}$ shortly before or after the inflaton $\phi$ decays into radiation $R$. Once the curvaton starts to oscillate, it redshifts like matter and comes to contribute a larger and larger fraction of the energy density, thus generating curvature fluctuations~\cite{Mollerach:1989hu,Mukhanov:1990me,Linde:1996gt,Moroi:2001ct,Lyth:2001nq,Moroi:2002rd,Lyth:2002my}.
	
	In general both the curvaton $\sigma$ and inflaton $\phi$ contribute to the curvature
	fluctuations on constant total density slicing $\zeta$. 	Depending on how dominant the curvaton is when it decays, as quantified by  
		\beq
	\rsig=\frac{\rho_\sigma}{(\rho_\sigma + 4\rho_R/3)} \Big|_D,
 	\eeq
	the relative contribution of inflaton and curvaton contributions to the total curvature varies, and is given by
	\beq
	\zeta=\zeta_\gamma=  (1-\rsig)\zeta_\phi + \rsig \zeta_\sigma,
	\eeq
where $\zeta_{i}$ is the curvature perturbation on constant density $\rho_i$ slicing
or equivalently the energy density perturbation $\delta \rho_i/3(\rho_i+p_i)$
on spatially flat slicing.  When applied to particle components, $\zeta_i$ is also
 the particle number density perturbation on spatially flat slicing.  Thus,~$i \in \{ \sigma,\phi, b,c,\nu,\gamma \}$.

   The curvaton can also generate isocurvature fluctuations \cite{Mollerach:1989hu,Linde:1996gt,Lyth:2001nq,Lyth:2003ip,Lemoine:2006sc}
   \begin{equation}
   S_{ij} = 3 (\zeta_i-\zeta_j),
   \end{equation}
   depending on how various particle numbers were generated. 
      If they were created before curvaton decay, then they inherit the
	inflaton's fluctuations $\zeta_\phi$. 
	If they were generated by curvaton decay, they inherit the curvaton's fluctuations
	$\zeta_{\sigma}$.
		If they were created from the thermal plasma after the curvaton decay, they inherit the total curvature perturbation $\zeta$.
	In summary~\cite{Lyth:2001nq,Lyth:2003ip},  \begin{equation}
    	 \zeta_{i} = 
	 \left\{
      	\begin{array}{ll}
          		\zeta_\phi, 	& \text{before  decay},\\
           	\zeta_\sigma , 	& \text{by decay},\\
            	\zeta, 						& \text{after  decay}.
         	\end{array}
     	\right.
 	 \end{equation}
Once generated, these curvature fluctuations remain constant outside the horizon~\cite{Bardeen:1980kt,Mukhanov:1990me,Malik:2004tf}.

We are interested, in particular, in the baryon ($b$) and
 cold dark matter ($c$) isocurvature 
fluctuations around the time of recombination.   We assume that lepton number is not related to curvaton
physics, allowing us to neglect neutrino isocurvature perturbations \cite{Lyth:2001nq,Lyth:2003ip}. We thus do not distinguish between photons $\gamma$ and the total radiation.

The remaining two isocurvature modes $S_{b\gamma}$ and $S_{c\gamma}$ can be 
reorganized into a CIP mode and a CDM isocurvature mode, called the effective
mode since it now carries all of the nonrelativistic matter isocurvature fluctuations, 
none of the baryon isocurvature fluctuations and only part of the CDM isocurvature
fluctuations.\footnote{This split between CIPs and an effective mode was introduced in Ref.~\cite{Gordon:2002gv} with the opposite convention of an effective baryon isocurvature,
rather than the CDM isocurvature mode.   The latter has since become standard (e.g.~\cite{Ade:2015lrj}).}
Specifically, we split each curvature fluctuation as
\begin{equation}
\zeta_i = \zeta_i^{\CIP}+  \zeta_i^{\eff} ,
\end{equation}
where by definition the CIP mode satisfies the compensation conditions $\delta \rho_b^\CIP= - \delta \rho_c^\CIP$ and $\delta\rho_\gamma^\CIP=0$ or equivalently
\begin{eqnarray}
\zeta_c^{\CIP} & {\equiv}&  - \frac{f_b}{1-f_b} \zeta_b^{\CIP} , \nonumber\\
\zeta_\gamma^{\CIP} & {\equiv}& 0,
\end{eqnarray}
and the effective mode carries the adiabatic fluctuations and CDM isocurvature fluctuations but  no baryon isocurvature fluctuations
\begin{equation}
\zeta_b^\eff  \equiv \zeta_\gamma^{\eff} = \zeta.\label{eq:eff_def_impl}
\end{equation} 

Here the baryon fraction is
	\begin{align}
	f_b = \frac{\rho_b}{\rho_b+\rho_c},
	\end{align}
	and we have assumed that the CIP mode is defined by compensation after both the
	baryons and CDM become nonrelativistic.   
	
We define the entropy perturbation carried by  the two modes as
\begin{equation}
S_{ij}^{X}=3(\zeta_{i}^{X}-\zeta_{j}^{X}),\end{equation} where $X\in \left\{\eff,{\rm CIP}\right\}$.
Equation~(\ref{eq:eff_def_impl}) then implies that the effective mode carries only CDM isocurvature fluctuations, as 
\begin{eqnarray}
S_{b\gamma}^\eff   &=& 0.
\end{eqnarray}
From these relations, we have
\begin{eqnarray}
S_{b\gamma}^\CIP &=& S_{b\gamma} ,\nonumber\\
S_{c\gamma}^\eff   &=& S_{c\gamma} + \frac{f_b}{1-f_b} S_{b\gamma}.
\end{eqnarray}
Together
$S_{c\gamma}^\eff$ and $S_{b\gamma}^\CIP$ give an alternate representation of
the isocurvature modes $S_{c\gamma}$ and $S_{b\gamma}$.
The benefit of this
representation is that because of the compensating baryon and CDM entropy fluctuations, the CIP mode corresponds to zero total isocurvature in nonrelativistic species [$S_{m\gamma}^{\rm CIP}\equiv f_{b}S_{b\gamma}^{\CIP}+(1-f_{b})S_{c\gamma}^{\CIP}=0$], and is unmeasurable in linear theory, while carrying all of the modulation of the baryon-photon ratio (since $S_{b\gamma}^{\rm eff}=0$), thus inducing potentially observable changes to CMB anisotropy properties at second order \cite{Holder:2009gd,Gordon:2009wx}.

Now let us consider the values of $S_{c\gamma}^\eff$ and $S_{b\gamma}^\CIP$ for the nine
baryon, CDM isocurvature scenarios, obtained by specifying whether or not the baryon number and CDM are set before, by, or after curvaton decay. We use the notation $(b_{x},c_{y})$, where $x,y\in\left\{{\rm before,~by, ~after}\right\}$, $b$ denotes baryon number, and $c$ denotes CDM.

Two curvaton-decay scenarios are of particular interest to CIPs.
 For the case when the baryon number is created by curvaton decay, $\zeta_{b} = \zeta_{\sigma}$ and CDM is created before, $\zeta_{c} = \zeta_{\phi}$,
 \begin{align}
		\frac{S_{c\gamma}^\eff }{\zeta_\sigma-\zeta_\phi}&= 3 \frac{ f_b - \rsig}{1- f_b} ,
\nonumber\\
		\frac{S_{b\gamma}^\CIP}{\zeta_\sigma-\zeta_\phi} &= 3(1-\rsig), \qquad (b_\by, c_\before),
\end{align}
and for baryon number created before and CDM by the decay,
\begin{align}
		\frac{S_{c\gamma}^\eff }{\zeta_\sigma-\zeta_\phi} &= 3\frac{1-f_b-\rsig}{1-f_b},
\nonumber\\
		\frac{S_{b\gamma}^\CIP}{\zeta_\sigma-\zeta_\phi} & = - 3 \rsig, 
		\qquad (b_\before, c_\by) .
\end{align}
In these two cases, $S_{c\gamma}^\eff$ can be made small by canceling the $\rsig$ and $f_b$ terms while leaving $S_{b\gamma}^\CIP$ relatively large.   
The other cases are given in Table~\ref{tab:curvatoncases}.  

 In all cases, the isocurvature modes are proportional to $\zeta_\sigma-\zeta_\phi$.
This implies that the cross-correlation between the curvature and isocurvature modes
share a universal correlation amplitude regardless of the curvaton-decay scenario.   For example, for the CIP mode
\begin{align}
R &\equiv
\frac{  P_{S_{b\gamma}^\CIP \zeta}}{\sqrt{ P_{\zeta \zeta} P_{S_{b\gamma}^\CIP S_{b\gamma}^\CIP}}}  \\
&= \pm
\frac{ (1-\rsig) P_{\zeta_\phi \zeta_\phi}-\rsig P_{\zeta_\sigma\zeta_\sigma}  } 
{\sqrt{  
(P_{\zeta_\phi\zeta_\phi}+ P_{\zeta_\sigma \zeta_\sigma}) [ (1-\rsig)^2 P_{\zeta_\phi \zeta_\phi} + \rsig^2 P_{\zeta_\sigma \zeta_\sigma  }]}} ,\nonumber
\end{align}
where
\begin{align} 
\langle \zeta_\sigma ^*(\bk) \zeta_\sigma(\bk')\rangle =& (2\pi)^3 \delta(\bk -\bk') P_{\zeta_\sigma \zeta_\sigma}(k), \nonumber\\
\langle \zeta_\phi^*(\bk) \zeta_\phi(\bk')\rangle = &(2\pi)^3 \delta(\bk -\bk') P_{\zeta_\phi \zeta_\phi}(k), \nonumber\\
\langle {S_{b\gamma}^\CIP}^*(\bk) \zeta(\bk')\rangle =& (2\pi)^3 \delta(\bk -\bk') P_{S_{b\gamma}^\CIP \zeta}(k), 
\end{align}
and we  have used the fact that the curvaton and inflaton fluctuations are uncorrelated,
\begin{equation}
P_{\zeta\zeta} = (1-\rsig)^2 P_{\zeta_\phi \zeta_\phi} + \rsig^2  P_{\zeta_\sigma \zeta_\sigma}.
\label{eqn:Pzetatotal}
\end{equation}
If either the curvaton or the inflaton dominates the total curvature, the CIP mode
is fully correlated ($R=\pm 1$), as is the CDM-isocurvature mode $S_{c\gamma}^\eff$.  The sign of the correlation depends on the decay scenario.

In fact, independently of curvaton domination or decay scenario, $S_{b\gamma}^\CIP$ and $S_{c\gamma}^\eff$ are 
always fully correlated and cannot be considered independently.   Whereas individually the CIP mode implies both a photon-baryon
 fluctuation and a CDM-baryon fluctuation, it can no longer be considered in isolation
 from the effective CDM isocurvature mode. This can lead to counterintuitive
 results when considering other representations of the isocurvature modes 
and determining their joint observational effects.

For example, in 
the $(b_\before,c_\before)$ case, both modes are present and in fact set the total $S_{bc}=0$. 
Obviously, this scenario cannot be tested through a spatial modulation of the baryon-CDM ratio. Nonetheless, the joint set of modes can be described by a CIP mode which carries 
$S_{bc}^{\CIP}$ and a fully correlated CDM isocurvature mode where $S_{bc}^\eff=-S_{bc}^\CIP$.   For the purposes of the tests in this paper, that is the more useful description, since CMB observables depend mainly on the photon-baryon modulation of quantities like the sound speed and damping scale of the plasma.   In this representation,
both the adiabatic and effective modes propagate in the presence of a CIP-modulated baryon-photon ratio.   Similarly, there are cases [($b_{\rm after},c_{\rm by})$ and ($b_{\rm after},c_{\rm before})$] where $S_{bc}\neq 0$, but the effective and adiabatic modes do not see a CIP-induced modulation of the baryon-photon ratio. Ultimately (as we see below),  the most interesting cases are those where the CIP mode is much larger than the effective mode, due to observational
bounds on the latter, making these subtleties largely irrelevant.

\setlength{\tabcolsep}{1.5em}
\begin{table*}[t]
\caption{CIP and CDM isocurvature modes for the various curvaton-decay scenarios.  
The center two columns give the general case, where both curvaton and inflaton fluctuations
contribute to the curvature fluctuations; whereas the right two columns give the curvaton-dominated
fluctuation case. Italicized cases are ruled out by observational bounds.  Bold-faced cases produce the largest CIP of the remaining ones and also predict anticorrelated
effective CDM isocurvature and curvature modes.}
\label{tab:curvatoncases}
\begin{ruledtabular}
\begin{tabular}{cc|cc|cc}
\rule{0pt}{4.5ex}
\rule[-3ex]{0pt}{0pt}
Baryons & CDM 
& $\dfrac{S_{c\gamma}^\eff}{\zeta_\sigma-\zeta_\phi}$ 
& $\dfrac{S_{b\gamma}^\CIP}{\zeta_\sigma-\zeta_\phi}$
& $\dfrac{S_{c\gamma}^\eff}{\zeta}$
& $A=\dfrac{S_{b\gamma}^\CIP}{\zeta}$ \\ \hline
\rule{0pt}{4.5ex}
\rule[-3ex]{0pt}{0pt}
{\bf by} & {\bf before} 
&  $- 3 \dfrac{  \rsig-f_b}{1- f_b}$ 
& $3(1-\rsig)$ 
& $-\dfrac{3}{\rsig} \dfrac{\rsig-f_b}{1-f_b}$ 
& $\dfrac{1-f_b}{f_b} \bigg( 3 + \dfrac{S_{c\gamma}^\eff}{\zeta} \bigg)$ 
\\
\rule{0pt}{4.5ex}
\rule[-3ex]{0pt}{0pt}
{\bf before} & {\bf by}
& $3\dfrac{1-f_b-\rsig}{1-f_b}$
& $- 3 \rsig$
& $ \dfrac{3}{\rsig} \dfrac{ 1 - f_b - \rsig}{1-f_b}$
& $ -3$
\\
\rule{0pt}{4.5ex}
\rule[-3ex]{0pt}{0pt}
by & after 
&  $3 f_b \dfrac{1-\rsig}{1-f_b}$
&  $3(1-\rsig) $
& $3\dfrac{ f_b }{\rsig} \dfrac{1-\rsig}{1-f_b} $
& $ \left( \dfrac{1}{f_b} -1 \right)\dfrac{S_{c\gamma}^\eff}{\zeta}$
\\
\rule{0pt}{4.5ex}
\rule[-3ex]{0pt}{0pt}
after & by 
&  $3 (1-\rsig)$
& $0$
& $3 \left( \dfrac{1}{\rsig}  - 1\right)$
& $0$
\\
\rule{0pt}{4.5ex}
\rule[-3ex]{0pt}{0pt}
{\it before} & {\it after}
& $ -3 \dfrac{f_b }{1-f_b}\rsig$
& $- 3 \rsig$
& $ -3 \dfrac{f_b }{1-f_b}$
& $-3$
\\
\rule{0pt}{4.5ex}
\rule[-3ex]{0pt}{0pt}
{\it after} & {\it before}
& $ -3\rsig$ 
& 0 
& $-3$
& $0$
\\
\rule{0pt}{4.5ex}
\rule[-3ex]{0pt}{0pt}
{\it before} & {\it before}
&  $-3 \dfrac{\rsig}{1-f_b}$
&  $- 3 \rsig$
& $ -3 \dfrac{1}{1-f_b}$
& $-3$
\\
\rule{0pt}{4.5ex}
\rule[-3ex]{0pt}{0pt}
by & by
&$ 3 \dfrac{1-\rsig}{1-f_b} $
& $3(1- \rsig)$
& $ \dfrac{3}{r_D} \dfrac{1-r_D}{1-f_b}$
& $(1-f_b) \dfrac{S_{c\gamma}^\eff}{\zeta}$
\rule{0pt}{4.5ex}
\rule[-3ex]{0pt}{0pt}
\\
after & after
& $0$
& $0$
& $0$
& $0$
\rule{0pt}{4.5ex}
\rule[-3ex]{0pt}{0pt}
\\
\end{tabular}
\end{ruledtabular}
\end{table*}

\subsection{Observational considerations for curvaton domination}\label{sec:cdom}
In forthcoming sections, we will consider the limit of fully correlated CIP modes ($R\approx\pm 1$), a case that
results if the curvaton completely dominates the total curvature fluctuation $\zeta \approx
\rsig \zeta_\sigma$.  
The inflaton fluctuations obey the usual relationship to tensor modes 
	\beq 
	\frac{P_{\zeta_\phi \zeta_\phi}}{P_{\zeta\zeta}} = \frac{r}{16\epsilon},\label{eq:inf_const}
	\eeq
where $r$ is the tensor-to-scalar ratio and $\epsilon$ is the slow-roll parameter from inflation. 
By comparing with Eq.~(\ref{eqn:Pzetatotal}), we see that the curvaton contribution to the total curvature $\zeta$ is dominant over the inflaton contribution as long as $ r \ll 16 \epsilon/ (1-\rsig)^2$.
Even if gravitational waves are detected near the current upper limit of $r \sim 0.1$,
 curvaton 
curvature domination can still be a good 
approximation for sufficiently large $\epsilon/(1- \rsig)^2$.   The remaining inflaton contribution
would then cause a small decorrelation of CIP modes which we ignore in the following 
sections.

Under the assumption that the inflaton fluctuations are negligible, there are tight constraints on the CDM-isocurvature fraction $S_{c\gamma}^\eff/\zeta$
   that then limit the CIP amplitude
   \begin{equation}
   A \equiv \frac{S_{b\gamma}^\CIP}{\zeta}
   \end{equation}
   in each of the nine scenarios. The two-sided 95\% CL constraints
   from the Planck 2015 temperature and low-$l$ polarization
    analysis of totally anticorrelated and correlated isocurvature
   modes with no tensors combine to imply \cite{Ade:2015lrj}
   \begin{equation}
   -0.080 <  \frac{S_{c\gamma}^\eff}{\zeta} < 0.042 \quad (\text{TT+lowP}).
   \end{equation}
   The asymmetric errors reflect the fact that there is a mild preference for anticorrelated
   CDM isocurvature modes in the Planck TT data \cite{Ade:2013uln}.  The standard adiabatic $\Lambda$CDM model
   predicts power in excess of the observations at low multipole moment which can 
   be canceled by such a mode. This preference would strengthen if existing upper limits to the amplitude of a primordial gravitational wave background are saturated in the future by a primordial $B$-mode detection \cite{Kawasaki:2014fwa}.  The preliminary Planck 2015 high-$l$ polarization, however, disfavors the anticorrelated scenario and leads to
   the bounds \cite{Ade:2015lrj}
      \begin{equation}
   -0.028 <  \frac{S_{c\gamma}^\eff}{\zeta} < 0.036 \quad (\text{TT,TE,EE+lowP})
   \end{equation}
   without tensors.
   
      Predictions for the various scenarios simplify in this curvaton-dominated limit.
   	The largest CIP amplitude is obtained if baryon number is created by the decay and CDM
	before
	\begin{align}
	\frac{S_{c\gamma}^\eff}{\zeta}  & = -\frac{3}{\rsig} \frac{\rsig-f_b}{1-f_b}, \nonumber\\
		A & =\frac{1-f_b}{f_b} \left( 3 + \frac{S_{c\gamma}^\eff}{\zeta} \right),  \quad (b_\by, c_\before).
\end{align}
Note that the CDM mode can be anticorrelated in this case but only satisfies observational
bounds if the decay fraction is tuned to near the baryon fraction $\rsig \approx f_b$.
The observational bound on $S_{c\gamma}^\eff/\zeta$ implies
$A$~$\approx$~$3(1$~$-f_b)/f_b$~$\approx 16.5$. 
The converse case gives
\begin{align}
		\frac{S_{c\gamma}^\eff}{\zeta} &= \frac{3}{\rsig} \frac{ 1 - f_b - \rsig}{1-f_b} ,
		\
\\
		A & = -3, \quad (b_\before, c_\by)\nonumber
\end{align}
and again allows anticorrelation and can satisfy observational bounds if $\rsig$ is
tuned to $1-f_b$.   

Table~\ref{tab:curvatoncases} lists the other cases.   For the $(b_{\rm by},c_{\rm after})$ and $(b_{\rm after},c_{\rm by})$
scenarios, $S_{c\gamma}^\eff/\zeta>0$, and to satisfy observational bounds, the CIP amplitude is 
either proportionately small {($|A|\sim 10^{-2}$)} or vanishing, {respectively}.
{The $(b_{\rm before},c_{\rm after})$ and $(b_{\rm after},c_{\rm before})$ scenarios cannot satisfy observational
bounds on $S_{c\gamma}^\eff/\zeta$  and are hence ruled out.   For the simultaneous scenarios,
$(b_{\rm before},c_{\rm before})$ cannot satisfy observational bounds, $(b_{\rm by}, c_{\rm by})$ predicts small CIP modes, and
$(b_{\rm after},c_{\rm after})$ predicts no isocurvature modes.}

In summary, the two cases that produce substantial CIP modes are the 
{$(b_{\rm by},c_{\rm before})$ and $(b_{\rm before},c_{\rm by})$
scenarios}, which predict $A \approx 3 (1-f_b)/f_b \approx 16.5$ and $A= -3$, 
respectively. Interestingly, these are also the only two scenarios where the CDM
isocurvature mode can cancel the excess low multipole power in the Planck TT data.

\section{CIP Reconstruction}\label{sec:reconstruction} 
CIPs leave observable imprints on the CMB. In this section, we review the method for CIP reconstruction introduced in Ref.~\cite{Grin:2011tf} and point out the limitations imposed by its use of a separate-universe approximation, as further illustrated in Appendix \ref{sec:wprop}. Reconstruction methods and results are general, and do not depend on whether or not CIPs are generated in a curvaton scenario.   Cross-correlations between the CIPs and the adiabatic mode do depend on the model.   Though the techniques again do not depend on the level of correlation, for results we assume fully correlated CIPs as appropriate for a curvaton-dominated scenario.

In Sec.~\ref{sec:SU}, we discuss the separate-universe response of off-diagonal short-wavelength CMB two-point correlations to the presence of a long-wavelength CIP mode.  This response is calculated by varying background cosmological parameters, as shown in Sec.~\ref{sec:calcresponse}.   Each two-point correlation function represents a noisy measurement of the CIP mode  which we combine to form the minimum variance estimator in Sec.~\ref{ssec:mve}.  We discuss its noise properties in
Sec.~\ref{sec:recnoise}. For correlated CIP modes motivated by the curvaton scenarios of Sec.~\ref{sec:curvaton}, the reconstruction can be correlated with the CMB fields themselves to enhance detection, as discussed further in Sec.~\ref{sec:sigma}.
				
\subsection{Separate-universe approximation}
\label{sec:SU}

The CIP mode $S_{b\gamma}^\CIP$ represents a modulation of the baryon-photon ratio
that is compensated by CDM so as to cancel its purely gravitational effects. Consequently, it  leaves no imprint on the CMB to linear order. At second order, other modes, including the dominant adiabatic mode,
propagate on a perturbed background where quantities such as the photon-baryon
sound speed and damping scale are spatially modulated.  A fixed 
CIP mode breaks statistical homogeneity and hence statistical isotropy in the CMB, and so the CIP can be reconstructed from the correlations between different 
CMB temperature and polarization multipoles that it induces.

In Ref.~\cite{Grin:2011tf}, an approximation for characterizing these off-diagonal
correlations was applied, based on what amounts to a separate-universe approximation \cite{Wands:2000dp}. Since the CIP
mode does not evolve, its impact can be characterized by a change in
cosmological parameters, so long as its wavelength is sufficiently large compared
with the scale over which the modes propagate.
In Appendix \ref{sec:wprop}, we show that for CMB anisotropy shortly after recombination, this requires the CIP mode to be
larger than the sound horizon at that time. This amounts to the limit $L\lsim 100$, where $L$ is the multipole index of the CIP projected onto the surface of last scattering during recombination. Use of the expressions derived here beyond this domain of validity will bias the associated CIP estimators, an issue we discuss further in Sec.~\ref{ssec:mve} and Appendix~\ref{sec:wprop}.

For modes that satisfy this approximation, we can treat the CIP mode shortly after recombination as a shift
in the background 
\begin{align} 
\Omega_b &\rightarrow  \Omega_b ( 1 + \Delta) ,\nonumber\\
\Omega_c & \rightarrow  \Omega_c - \Omega_b \Delta,
\label{eqn:separate}
\end{align}
where
\begin{equation}
\Delta(\bn)  = S_{b\gamma}^\CIP(\bx=D_*\bn), 
\label{eqn:angularCIP}
\end{equation}
 $\bn$ is the direction on the sky, and $D_*$ is the distance to the CMB last-scattering surface during recombination.
This angular field can be decomposed into multipole moments 
		\beq
			\Delta(\bn) = \sum_{LM} 	\Delta_{LM} 	Y_{LM}(\bn),\label{eq:proj}
 		\eeq
so that the restriction on the wavelength of the CIP may be considered as a low-pass
filter where the effects are characterized out to $L \lesssim 100$.  Note that this
restriction justifies the use of a single distance in Eq.~(\ref{eqn:angularCIP}) rather than
an average over the finite width of the recombination era. 

In linear theory,
the impact of background parameters 
on CMB power spectra 
\begin{equation}
			\langle X^{*}_{l'm'}Z_{lm} \rangle = 
			\delta_{l l'}\delta_{mm'}C_{l}^{XZ},
			\label{eqn:twoptdiag}
\end{equation}
are characterized
by transfer functions
\begin{align}
			C_l^{XZ} = \frac{2}{\pi}\int k^{2}dk T_{l}^X(k) T_l^{Z}(k)P_{\zeta\zeta}(k)
			\label{eqn:Cl}
\end{align}
that are given by integral solutions to the Einstein-Boltzmann equations of radiative
transfer.   Here  $X$ and $Z$ are any of the CMB temperature and polarization fields $T,E,B$.
Given observational bounds on the CDM isocurvature mode, to a good approximation 
we can set $S_{c\gamma}^\eff=0$  when evaluating the transfer functions in Eq.~(\ref{eqn:Cl}). In the curvaton model with $S_{b\gamma}^\CIP= A \zeta$, the $\Delta$ field is
 correlated with the CMB fields through their joint dependence on $\zeta$.  For the case where the curvaton dominates $\zeta$ and in a flat
 cosmology,  the cross power-spectrum of $\Delta$ and the
CMB fields $C_l^{X\Delta}$ as well as the auto power-spectrum  $C_l^{\Delta\Delta}$  are described by Eq.~(\ref{eqn:Cl}) with
\begin{equation}
T_l^{\Delta}(k) =A j_l(k D_*)
\end{equation}
and can be numerically evaluated in $\textsc{camb}$~\cite{Lewis:2002nc} using Eq.~(\ref{eqn:Cl}).

Note that even in the presence of CIP modes, which are themselves 
statistically isotropic, 
two-point correlations are characterized by the diagonal form of Eq.~(\ref{eqn:twoptdiag}) as long as 
$\langle \ldots \rangle$ is understood to be the ensemble average over realizations of
all modes.   In fact, in the curvaton model, the ensemble average over realizations of
$\zeta$ automatically includes the CIP and adiabatic modes.

Nonetheless, it is useful to artificially separate the two and consider the response of
CMB fields to a fixed realization of the CIP mode.   
  In the SU 
 approximation, this fixed CIP  mode is treated as simply a change in cosmological parameters
 from Eq.~(\ref{eqn:separate}) that varies across the sky~\cite{Grin:2013uya}.  
 This variation modulates the statistical properties of the CMB modes according
 to the transfer functions.
 The utility of this split is that it exposes the fact that there are many pairs of CMB multipoles 
where $l,l' \gg L$ that can be used to estimate the realization of $\Delta_{LM}$ on our sky.
 We denote an average over CMB modes with the CIP mode fixed as
 $\langle \ldots \rangle_\CMBavg$.  This average can be thought of as an average over the subset of $\zeta$ modes that are smaller in wavelength than the sound horizon in the presence of fixed
 longer-wavelength $\zeta$ modes.

\begin{table}
\caption{The response function $S^{L, XZ}_{l l'}$ of the various two-point observables
 in Eq.~(\ref{eq:response}).	} 
\begin{ruledtabular}\label{tab:response} 
\begin{tabular}{c c c} 
$XZ$		& $S^{L, XZ}_{l l'}$ & $l + l' + L$\\	\noalign{\smallskip}
\hline \noalign{\smallskip}
$TT$		& $(C^{T,dT}_{l'} + C^{T,dT}_{l}) K^{L}_{ll'}$& even\\ 	\noalign{\smallskip}
$EE$ 		& $(C^{E,dE}_{l'} + C^{E,dE}_{l}) H^{L}_{ll'}$& even \\ 	\noalign{\smallskip}
$EB$		& $-i (C^{E,dE}_{l'} + C^{B,dB}_{l})H^{L}_{ll'}$ & odd \\	\noalign{\smallskip}
$TB$		& $-i C^{T,dE}_{l'} H^{L}_{ll'}$ & odd	 \\ \noalign{\smallskip}
$TE$ 		& $(C^{T,dE}_{l'} H^{L}_{ll'} + C^{E,dT}_{l} K^{L}_{ll'}$) & even \\ \noalign{\smallskip}
$BB$		& $(C^{B,dB}_{l'} + C^{B,dB}_{l}) H^{L}_{ll'}$& even\\ 
\end{tabular} 
\end{ruledtabular}
\end{table}

The product of the source and modulation fields in real space
leads to a convolution in harmonic space. Hence, it connects
CMB multipole moments of different $l,m$ in the same manner as a three-point function, yielding 
 		\begin{align} \label{eq:response}
     			\langle X^{*}_{l'm'}Z_{lm} \rangle_\CMBavg = & \,	\delta_{l l'} \delta_{m m'}	C_{l}^{XZ}  
\notag\\ 
			& +  \sum_{LM} \Delta_{LM}	\xi^{LM}_{lm, l' m'} 	S^{L, XZ}_{l l'},
		\end{align}	
 where 
 \begin{align}		
    			 \xi^{LM}_{l m l' m'} 		
			= & \,	(-1)^{m}		\sqrt{\frac{(2L+1)(2l+1)(2l'+1)}{4\pi}}	
\notag \\			
			& \times 	     \wigner{l}{-m}{L}{M}{l'}{m'},							\end{align}		
and the response functions $S^{L,XZ}_{ll'}$ are given in Table~\ref{tab:response} with
\begin{align}
			C_l^{X,dZ} = \frac{2}{\pi}\int k^{2}dk T_{l}^X(k) \frac{d T_l^{Z}}{d\Delta}(k)P_{\zeta\zeta}(k)
			\label{eqn:Clderiv}
\end{align}
and
\begin{align}
    			K^{L}_{ll'}	\equiv &		\wigner{l}{0}{L}{0}{l'}{0}, \nonumber\\
		    	H^{L}_{ll'}	\equiv &		\wigner{l}{2}{L}{0}{l'}{-2}, \nonumber
\end{align}
which are Wigner 3$j$ coefficients.
The response for intrinsic $B$ modes is new to this work and may provide
extra information on CIP modes should they be detected in the future.

In the presence of a  fixed long-wavelength CIP mode, statistical isotropy of short-wavelength
CMB fields is therefore broken.  Statistical isotropy is of course restored
once the full ensemble average over the random realizations of the CIP mode is taken.  
Given the correlation of $\Delta$, $\zeta$ and the $T,E$ CMB fields, 
a full ensemble average
induces a three-point correlation in the CMB which correlates long-wavelength modes to short-wavelength power, i.e. a squeezed bispectrum.   

\begin{figure}
\includegraphics[scale=0.60]{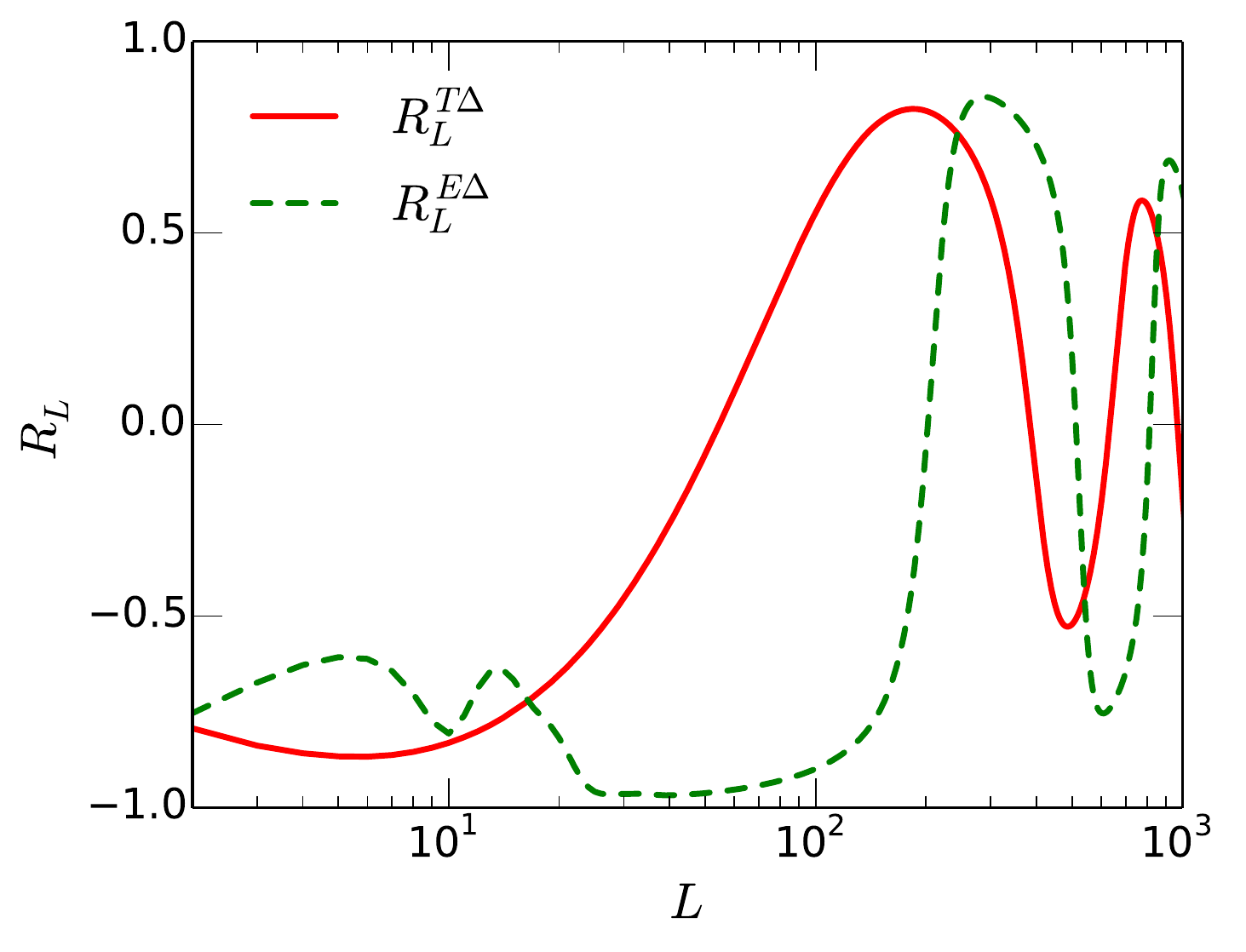}
\caption{\label{fig:R} 
Correlation coefficients $R_L^{T\Delta}$ and $R_L^{E\Delta}$ between the CIP and the CMB temperature and polarization fields, respectively, as a function of multipole $L$ for $A>0$. The sign of the correlation 
oscillates due to acoustic oscillations in temperature and polarization, whereas the level
of correlation depends on the difference in projection effects between the fields.  }
\end{figure}

 This correlation 
provides
a way of detecting the CIP mode from a noisy two-point reconstruction of $\Delta$
as long as  
the correlation coefficient 
\beq 
	R_L^{X\Delta} = \frac{C_L^{X\Delta}}{\sqrt{C_L^{XX} C\DD}}\label{eq:rxdef},
\eeq
 shown in Fig.~\ref{fig:R}, remains large. The sign of the correlation 
oscillates due to acoustic oscillations in temperature and polarization, whereas the level
of correlation depends on the difference in projection effects between the fields. 

\subsection{Response of CMB anisotropies to CIP modes}
\label{sec:calcresponse}

Calculating the response functions in Eq.~(\ref{eqn:Clderiv}) requires 
varying cosmological parameters to mimic the effect of the CIP within the separate-
universe approximation following Ref.~\cite{Grin:2011tf}.   While Eq.~(\ref{eqn:separate}) provides a prescription for
the main effect of compensating variations of the background baryon and cold dark matter densities, there are a number of subtleties that arise from the treatment of the CIP mode as
an angular field  $\Delta(\bn)$ rather than a three-dimensional field that
varies along the line of sight.  CMB temperature and polarization anisotropies that are generated
at reionization (and thus after recombination) break this approximation. For these sources of CMB anisotropies, the use of
the SU approximation is limited by the horizon scale at the given
time rather than the sound horizon shortly after recombination.

By varying $\Omega_b$ and
 $\Omega_c$ in the transfer functions, we implicitly include a reionization response to the CIP that depends on what other parameters
are held fixed.   Even if we assume that the reionization optical depth $\tau$ is held fixed when 
varying parameters in the transfer function, there is still an effect on the shape of the polarization spectra due to the implied modulation of 
the redshift of reionization $z_r$.   To assess the impact of reionization, we try two other prescriptions that attempt to remove
this sensitivity.   The first case is to simply adopt a model with no reionization, and the 
second is to neglect reionization in evaluating  Eq.~(\ref{eqn:Clderiv}) and then restore it
 using the analytic damping envelope of Ref.~ \cite{Hu:1996mn}. All  three  prescriptions yield similar (at the $10\%$$-$$20\%$ level) results for the sensitivity of the CMB to correlated CIPs in Sec.~ \ref{sec:sigma}.  
We conclude that that reionization only causes a small ambiguity for the detectability of CIPs.
If they are in the future detected and measured precisely, then a more detailed prescription
will be required.   
For simplicity, we adopt here the constant $\tau$ prescription.  

Similarly, the CMB fields from shortly after recombination are gravitationally lensed by large-scale structure in the foreground.   
Gravitational lensing of the CMB also produces
off-diagonal two-point correlations in the presence of a fixed large-scale lensing potential.
Given differences in the response function, it is in principle possible to disentangle 
lensing from CIP effects internally to the CMB.   Likewise, external delensing of the CMB
can help remove the contamination.   These topics will need to be addressed in the future
but are beyond the scope of the present work.   They
 will  degrade somewhat the forecasts for CIP detectability due to loss of degenerate modes.   
 In the following sections, we treat gravitational
lensing effects as a source of additional Gaussian noise only and continue to use $C_l$ to 
denote the unlensed CMB power spectrum.

Throughout this work, we use a flat $\Lambda$CDM cosmology consistent with the 2013 Planck results \cite{Ade:2013zuv}.\footnote{Specifically, we use baseline model 2.13 from \href{http://wiki.cosmos.esa.int/planckpla/images/9/9b/Grid_limit68.pdf}{Grid\_limit68}.
}   
Here $\Omega_b = 0.049$ and $\Omega_c = 0.268$, around which we calculate the CIP response, and the adiabatic scalar power spectrum with amplitude $A_s = 2.215\times10^{-9}$, spectral index $n_s = 0.9624$, the reionization optical depth $\tau = 0.0925$, neutrino mass of a single species $m_{\nu}= 0.06$ $\mathrm{eV}$, and Hubble constant $h$~=~0.6711, which we hold fixed.  We assume that the tensor modes (parameterized by the tensor-to-scalar ratio $r$) are negligible, and thus that there are no intrinsic $B$ modes.

\subsection{Minimum-variance CIP estimator}\label{ssec:mve}

Each pair of CMB fields $ X^{*}_{l'm'}Z_{lm}$ provides an estimate of a CIP
mode $\Delta_{LM}$ that satisfies the triangle inequality,  $ | l-l'| \le L \le l +l '$ and $M = m-m'$.  
Any single pair, however, is highly noisy due to the sample variance of the Gaussian random
CMB fields and instrument noise. Here, we also include the change in the power spectra due to lensing as an additional noise source. 
Including all noise sources means that we replace
Eq.~(\ref{eqn:twoptdiag}) with
\begin{equation}
			\langle X^{*}_{l'm'}Z_{lm} \rangle = 
			\delta_{l l'}\delta_{mm'}\tilde C_{l}^{XZ},
\end{equation}
where 
\bea \label{eq:Cl_def}
\tilde{C}_l^{XZ} &=& C_l^{XZ} + \delta C_l^{XZ,{\rm lens}} + \tilde N_l^{XZ},
\eea
with $\tilde N_l^{XZ}$ as the measurement noise power in the sky maps and $\delta C_l^{XZ,{\rm lens}}$ as the change in the power spectrum due to lensing, which we treat as noise.
Furthermore, 
given a $XZ$ field pairing, the $X^{*}_{l'm'}Z_{lm}$ and $Z^{*}_{l'm'}X_{lm}$ estimators
have correlated noise.   Likewise, the different field pairings $XZ$ and $X'Z'$ are also
correlated.  Following the mathematically identical lensing treatment in Ref.~
\cite{Okamoto:2003zw} (see also Ref.~\cite{Namikawa:2011cs}), we optimize the weighting of
the estimators to minimize the CMB reconstruction noise.    Optimal weighting generalizes
the familiar inverse-variance weighting to inverse-covariance
weighting for covarying estimators.  Differences with the results of Ref.~\cite{Grin:2011tf}
are discussed in Appendix~\ref{sec:correct_noise}.

It is convenient to break the inverse-covariance weighting into two steps.   Since we want to
examine each $XZ$ pair individually, we first
consider the $2\times 2$ covariance of its multipole pairing.  
We can write a general estimator as
\begin{equation}\hat{\Delta}_{LM}^{XZ}=\sum_{lm l'm'} X^{*}_{l'm'}Z_{lm}\xi^{LM}_{lm l'm'}W_{L l l'}^{XZ},\label{eq:est_form}
\end{equation}
where $\xi^{LM}_{lm l'm'}$ enforces the triangle inequality and $W_{L l l'}^{XZ}$
are the weights to be determined. 
The variance of an unbiased estimator due to Gaussian CMB noise  becomes
\begin{widetext}\begin{align}
\left \langle |\hat{\Delta}_{LM}^{XZ}-\Delta_{LM}|^{2}\right \rangle_{\rm CMB}=\sum_{l l'}&G_{l l'}W_{L l l'}^{XZ}\left\{\tilde{C}_{l'}^{XX}\tilde{C}_{l}^{ZZ}W_{L l l'}^{XZ*} +(-1)^{l+l'+L}\tilde{C}_{l'}^{XZ}\tilde{C}_{l}^{XZ}W_{Ll' l}^{XZ*}\right\},\label{eq:variance_general}
\end{align}
where we have used Wick contractions and identities for the Wigner coefficients, and \begin{equation}
G_{l l'}\equiv \frac{(2l+1)(2l'+1)}{4\pi}.
\end{equation}
Note that the covariance of the multipole permutation gives rise to the second term.
We minimize the variance by taking a derivative of Eq.~(\ref{eq:variance_general}) with respect to $W_{L l l'}^{XZ}$, imposing the constraint of an unbiased estimator with a Lagrange multiplier $\lambda$, obtaining
\begin{align}
\tilde{C}_{l'}^{XX}\tilde{C}_{l}^{ZZ}W_{Ll l'}^{XZ*}+\left(-1\right)^{l+l'+L}\tilde{C}_{l'}^{XZ}\tilde{C}_{l}^{XZ}W_{Ll' l}^{XZ*}+\lambda S_{l l'}^{L,XZ}=0,\label{eq:con_a} 
\end{align}
This provides the relative weights $W_{Ll l'}^{XZ*} \propto g_{Ll l'}^{XZ*}$,
where 
\begin{align}
g_{L l l'}^{XZ}=\frac{S_{l l'}^{L,XZ*}\tilde{C}_{l}^{XX}\tilde{C}_{l'}^{ZZ}-\left(-1\right)^{l+l'+L}S_{l' l}^{L,XZ*}\tilde{C}_{l}^{XZ}\tilde{C}_{l'}^{XZ}}{\tilde{C}_{l'}^{XX}\tilde{C}_{l}^{ZZ}\tilde{C}_{l}^{XX}\tilde{C}_{l'}^{ZZ}-\left(\tilde{C}_{l}^{XZ}\tilde{C}_{l'}^{XZ}\right)^{2}}.
\label{eq:single_noise}
\end{align}
Note that this takes the form of  the inverse $2\times 2$ covariance weight as expected.

The normalization comes from the requirement that the estimator be unbiased so that
\begin{align}
\hat{\Delta}_{LM}^{XZ}&= N_{L}^{XZ} \sum_{lm l'm'}X_{l'm'}^{*}Z_{lm}g_{L l l'}^{XZ}\xi^{LM}_{lm l'm'},\label{eq:single_estimator}\\
\left[N_{L}^{XZ}\right]^{-1}&=\sum_{l l'}G_{l l'}S_{l l'}^{L,XZ}g_{L l l'}^{ XZ}.\label{eq:single_norm}\end{align}
The normalization factor is also the variance of the 
estimator itself,
\begin{equation}
\left \langle |\hat{\Delta}_{LM}^{XZ}-\Delta_{LM}|^{2} \right \rangle _{\mathrm{CMB}}=N_{L}^{XZ}.\label{eq:var}
\end{equation}

Next, we can combine the various $\alpha =XZ$ field pairs to find the 
total minimum-variance estimator by again inverse-covariance weighting the
individual estimators
\bea
\hat{\Delta}_{LM}&=&\sum_{\alpha} w_L^{\alpha}\hat{\Delta}_{LM}^{\alpha},\qquad
w^{\alpha}_L=N_{L}^{\Delta\Delta}\sum_{\beta}\left( \mathcal{M}_{L}^{-1}\right)^{\alpha,\beta},\qquad
 \left[N_{L}^{\Delta \Delta}\right]^{-1} \equiv{\sum_{\alpha\beta}\left(\mathcal{M}_{L}^{-1}\right)^{\alpha,\beta}}. \label{eq:full_estimator}
\eea The estimator covariance-matrix $\mathcal{M}_{L}$ is at every $L$ a rank-$2$ tensor over observable pairs. The indices $\alpha$ and $\beta$ take values over labels for pairs of observables, that is, $\alpha,\beta\in \left\{TT,EE,TE,BT,BE\right\}$. Using 
Eq.~(\ref{eq:single_estimator}), and identities of Wigner coefficients, we obtain an expression for the matrix elements $\mathcal{M}_{L}^{\alpha,\beta}$:\begin{eqnarray}
\mathcal{M}_{L}^{XZ, X'Z'}=N_{L}^{XZ}N_{L}^{X'Z'}\sum_{l l'}G_{l l'}g_{L l l'}^{XZ}\left[ \tilde{C}_{l'}^{XX'}\tilde{C}_{l}^{ZZ'}g_{L l l'}^{X'Z'*}+\left(-1\right)^{l+l'+L}\tilde{C}_{l'}^{XZ'}\tilde{C}_{l}^{X'Z}g_{L l' l}^{X'Z'*} \right]. \label{eq:covmat_total}
\end{eqnarray}
The total-estimator variance is again the normalization factor $N_{L}^{\Delta \Delta}$.  
It is straightforward to check that Eqs.~(\ref{eq:single_norm}) and (\ref{eq:var}) can be recovered from Eq.~(\ref{eq:covmat_total}) by restriction to a single pair ($X=X'$, $Z=Z'$). 

\end{widetext}

\setlength{\tabcolsep}{0em}
\begin{table}[t]	
\caption{Instrument noise parameters for illustrative experiments~\cite{wmapnoise,Planck:2013cta,2010SPIE.7741E..1SN,sievers,2013ApJ...765L..32K,crawford,Abazajian:2013oma}: full-width half-max (FWHM) of the beam (in arcmin), noise for temperature measurements $\delta_{TT}$ (in $\mu$K$~{\rm arcmin}$) and, sky fraction $f_{\rm sky}$ used for cosmological analysis.  For Planck, we indicate temperature and polarization noise separately as described in the text.   For reconstruction, we minimum variance weight the V and W bands for WMAP and  the 143 and 217~GHz channels for Planck.
 The CVL case is full sky and has no instrument noise by definition.}
\begin{ruledtabular}
\begin{tabular}{llll}
Data&FWHM & Noise&$f_{\rm sky}$ 	
\\ \hline 	\noalign{\smallskip}
WMAP  V band & 21   &434&0.65 \\ 
WMAP   W band& 13  & 409 &0.65\\  
Planck 143~{\rm GHz}& 7.1  & 37, 78&0.65\\  
Planck   217~{\rm GHz}& 5.0  & 54, 119&0.65\\  
ACTPol&1.4&8.9&0.097\\  
SPT-3G&1.1&2.5&0.06\\ 
CMB-S4&3.0&1.0&0.50\\  
CVL&0.0& 0.0&1\\     
\end{tabular}
\label{tab:instrument_noise} 
\end{ruledtabular}
\end{table}
\begin{figure}[!h]
\includegraphics[width=3.5 in ]{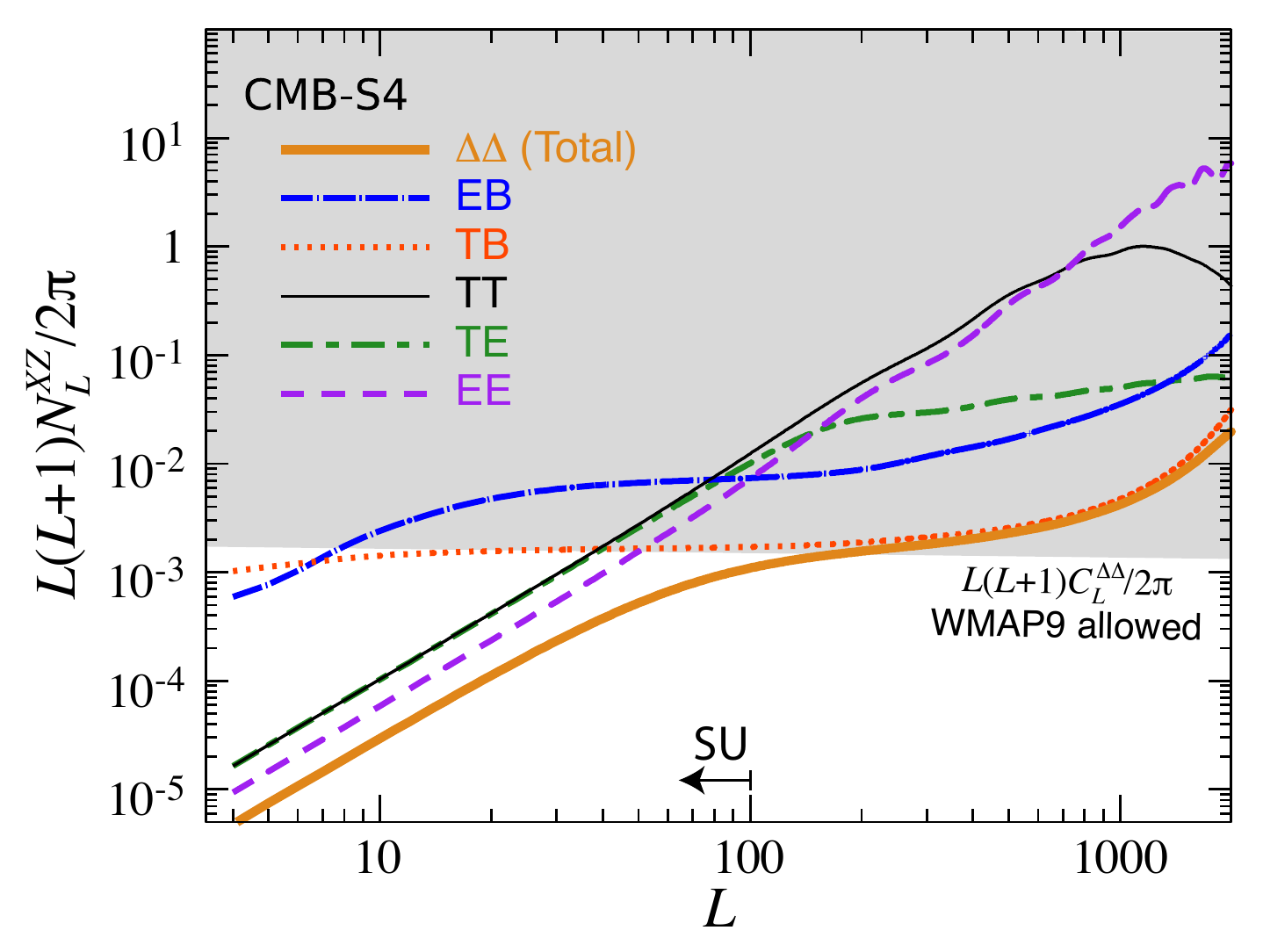}
\caption{Reconstruction noise curves $N_{L}^{XZ}$ for CMB-S4 obtained from the CIP estimator of Eq.~(\ref{eq:single_estimator})
with the indicated pair of observables $XZ$.  The shaded region represents $L(L+1)C_{L}^{\Delta \Delta}/(2\pi)$ signals that are excluded by WMAP 9-year data (Ref.~\cite{Grin:2013uya}, see Sec.~\ref{sec:sigma}).  This bound comes from limits to the auto-correlation power spectrum $C_{L}^{\Delta \Delta}$ of CIPs and is thus conservative for models with correlated CIPs. Where the curves intersect this bound, the estimator noise and CIP sample variance  of the $|A|=808$ model that saturates it are
equal.  The approximate domain of validity of the SU approximation $L \lesssim 100$ is
indicated by the arrow. Unless otherwise specified we assume, the estimators employ 
CMB multipoles up to $l_{\rm CMB}=2500$ throughout. }
\label{fig:recon_noise}
\label{fig:nc}
\end{figure}

\subsection{Reconstruction noise}
\label{sec:recnoise}
For reconstruction noise forecasts $N_L^{XZ}$ from the CMB fields $XZ$, we can use the $\tilde N_l^{XZ}$ CMB noise power specifications (real or projected) of various experiments.   We parameterize it as
\bea
\tilde N_{l}^{XZ} &=& \delta_{XZ}^2 e^{l\left(l+1\right)\theta_{\rm FWHM}^{2}/8\ln 2},\label{eq:instrument_noise}
\eea
with
$\delta_{XZ}^2$ as the detector noise covariance
assumed to be zero if  $X \neq Z$,
and $\theta_{\rm FWHM}$ as the full-width half-max of an approximately Gaussian beam.
Table~\ref{tab:instrument_noise} gives the specifications for WMAP \cite{wmapnoise}, {Planck} \cite{Planck:2013cta}, ACTPol \cite{2010SPIE.7741E..1SN,sievers}, SPT-3G \cite{2013ApJ...765L..32K,crawford}, and CMB Stage 4 (CMB-S4 henceforth) \cite{Abazajian:2013oma} experiments. 
The WMAP and {Planck} missions have concluded, but because WMAP data have not yet been used to search for correlated CIPs, and because no CIP reconstruction yet exists from {Planck} data, we "predict"  in those cases as well. 
  For all but the Planck case,
we take $\delta_{EE}^2=\delta_{BB}^2 = 2 \delta_{TT}^2$. 
For Planck, not all HFI bolometers have polarization sensitivity. To forecast temperature noise, we use all $3S + 4P$ and $4S + 4P$ bolometers from the 143 and 217~GHz channels respectively, where $S$ denotes an unpolarized spider-web bolometer and $P$ a polarized bolometer.
We also calculate the ideal reconstruction noise for the (zero instrument-noise) hypothetical cosmic-variance-limited (CVL) case as the ultimate limit.  

The results for CMB-S4 are shown in Fig.~\ref{fig:recon_noise} (see Appendix~\ref{sec:correct_noise} for other experiments).
For all cases, we generate reconstruction noise curves using a maximum observed CMB multipole index $l_{{\rm CMB}}=2500$.   Beyond this point, foregrounds dominate the TT
spectrum.
To assess the possible effect of CMB foreground subtraction and lower contamination in the polarization spectra, we explore 
increasing the limit in Sec.~\ref{sec:sigma}. Also shown in Fig.~ \ref{fig:recon_noise} is the bound of $|A|<808$ on the CIP signal power converted from Ref.~\cite{Grin:2013uya} for the fiducial cosmology.  This bound comes from limits to the auto-correlation power spectrum $C_{L}^{\Delta \Delta}$ of CIPs. It is thus valid but could be further improved if $\Delta$ and the primordial curvature $\zeta$ are correlated, as we discuss further in Sec. \ref{sec:sigma}.

The CIP estimator used in all these forecasts, Eq.~(\ref{eq:single_estimator}), is derived under the SU approximation. Naive application of Eq.~(\ref{eq:response}) outside its regime of validity can thus significantly bias estimates of the CIP amplitude $\Delta_{LM}$. In the context of the simple toy model in Appendix~\ref{sec:wprop}, we can compute the response exactly and estimate 
the reduced sensitivity and bias beyond the SU approximation.
Using those results, we estimate that the reconstruction noise curves
are accurate for  $L\lsim 100$.  Moreover, the sharply reduced response beyond this point in the toy model implies
that we lose little information by simply restricting ourselves to these $L$ values.

The CIP constraints of Ref.~\cite{Grin:2013uya} are unaffected by the limitations of the SU approximation, as the WMAP signal-to-noise ratio (S/N) for CIPs is dominated by scales $L\lsim 10$. Planck will also not run up against the limitations of the SU estimator. On the other hand,  future experiments will not be as sensitive to CIPs as naively estimated in the SU approach, as additional information from polarization-based estimators is severely reduced when the $L\lsim 100$ limit is imposed. 
We discuss further the limitations to the detection of curvaton-generated CIPs in Sec.~ \ref{sec:sigma}. 

It is instructive to develop some intuition for the shapes of the reconstruction-noise curves in Fig.~\ref{fig:recon_noise}. From the geometry of converting $E$ into $B$ alone, we can understand the relative slopes of $N_{L}$ between $B$-based and non-$B$-based estimators of $\Delta_{LM}$.  The difference is easiest to see in the flat-sky approximation.  
In order to generate a $B$ mode from the CIP modulation of an 
$E$ mode, the modulation must change the direction of the mode relative to the polarization direction. In the squeezed limit
where $l,l' \gg L$, the flat-sky correspondence gives Fourier modes where $\bl \parallel 
\bl'$ and so does not generate a $B$ mode.  In the full-sky formulas this comes about
because to good approximation~\cite{Hu:2000ee}
\begin{eqnarray}
H_{l l'}^{L}&\approx & F(l,L,l')
\begin{cases}
\sin (2\phi), & l+l'+L~{\rm odd}, \\
\cos (2\phi), &  l+l'+L~{\rm even} ,
\end{cases}
\end{eqnarray}
where $\phi$ is the angle between the $l$ and $l'$ sides of a triangle with side
lengths $\{ l,L,l' \}$  upon which $F$ also depends.
In the squeezed limit, $\sin(2\phi) \propto L$ and $\cos(2\phi) \approx 1$, which explains 
why $N^{XB}_{L}/N_{L}^{XE}\propto L^{-2}$, as we can see is the case in Fig.~\ref{fig:recon_noise}. 
We also see in Fig.~\ref{fig:recon_noise} that for $L \lesssim 100$, the noise spectra of estimators that do not involve $B$ are white.  This  behavior is to be expected, given that the prominent
features in the CMB carry the scale of the acoustic peaks so that the noise for a modulation by a smaller $L$ does not depend on $L$. Finally, $N_L^{TE}$ changes slope at roughly the sound horizon scale $L \approx 100$. This is because of the acoustic phase difference between $T$ and $E$.   The two terms in the response $C_l^{T,dE}$ and $C_l^{E,dT}$ are opposite in sign and roughly out of phase by half a period $\Delta l \approx 100$. So for $l$, $l'$ that differ by less than half a period, the response  remains small, while for $l$, $l'$ separated by more than a period, the response grows.  Note, however, that the  $L \gtrsim 100$ required for the latter
is the region for which the SU approximation fails, so 
this enhancement of the response is not of practical value.

\section{Correlated CIP Forecasts}\label{sec:sigma}
We now forecast constraints on the amplitude $A$ of totally correlated (or anticorrelated) CIP modes for the various experiments in 
Table~\ref{tab:instrument_noise}. We thus restrict ourselves to the curvaton-dominated limit described in Sec.~\ref{sec:cdom}. We use Fisher information matrix techniques in Sec.~\ref{sec:fisher}, and then discuss the dependence of these results
on various aspects of the data and assumptions in Sec.~\ref{sec:dependencies}.

\subsection{Fisher errors}
\label{sec:fisher}
The Fisher information matrix $F_{ij}$ forecasts the inverse-covariance matrix of a set of parameters $p_i$, including $A$, on which
the auto- and cross-spectra pairs $\alpha=XZ$ of the  observed CMB and reconstructed CIP fields, $\{ X,Z \} \in  \{T,E,B,\Delta\}$ depend.   
Under the assumption of Gaussian statistics for these underlying fields,  the Fisher information matrix can be approximated as
\beq 
	F_{ij}  = \sum_{L_{\mathrm{min}}}^{L_{\mathrm{max}}} 	(2L+1) \fsky \sum_{\alpha,\beta} 
	\frac{\partial C_L^{\alpha}}{\partial p_i}	\left({\cal C}^{-1}_{L}
	\right)^{\alpha\beta} \frac{\partial C_L^{\beta}}{\partial p_j},
	\label{eq:fisher}
\eeq
where ${\cal C}_L$ is the covariance matrix for an individual $LM$ mode in the power spectrum estimator
\beq
			{\cal C}^{XX',ZZ'}_{L} ={ \tilde{C}_L^{XZ} \tilde{C}_{L}^{X'Z'} + \tilde{C}_L^{XZ'} \tilde{C}_{L}^{X'Z}},
\eeq
and the $\fsky$ factor roughly accounts for the reduction in the $2L+1$ independent $M$ modes due to the sky cut.
Unless otherwise specified, we always employ the reconstruction-noise power spectrum $N\DD$ for a $\Delta$ reconstructed from  CMB multipoles up to
$l_{\rm{CMB}} = 2500$. 

\begin{table}[htbp]	
\caption{ $2\sigma_A$ detection threshold for $A$ given all auto- and cross-spectra (see Sec.~\ref{sec:fisher} for assumptions).  Cross-spectra allow considerable improvement from the current bounds, and CMB-S4 is able to probe the largest prediction for the curvaton model ($A\approx16.5$) at more than $3\sigma$ significance. 
}
\label{tab:2sigma} 
\setlength{\tabcolsep}{1.5em}
\begin{ruledtabular}
\begin{center}
\begin{tabular}{ll}
          Data & $2\sigma_A$ \\ \hline     	\noalign{\smallskip}
WMAP    & 152      \\ 
Planck   & 43.3           \\ 
ACTPol   & 40.2          \\ 
SPT-3G   & 38.2       \\ 
CMB-S4   & 10.3            \\ 
CVL      & 6.5         
\end{tabular}
\end{center}
\end{ruledtabular}
\end{table}	

In our case, we are interested in the parameter $p_i= A$.  Information on $A$ is contained in the auto-spectrum
of the reconstruction $C_L^{\Delta\Delta}\propto A^2$, the cross-spectra with the CMB $T$ and $E$ fields $C_L^{T\Delta}$, 
$C_L^{E\Delta} \propto A$, and in principle $C_L^{BB}$. We choose to neglect the information coming from $BB$, because CIP $B$-mode power will be swamped by the lensing spectrum by a factor of $\sim 10^{2}$ for the WMAP9 allowed model shown in Fig.~\ref{fig:recon_noise}~\cite{Grin:2011tf}.  

Since other cosmological parameters that define the curvature power spectrum $\Delta_\zeta^2$ and matter content
are well determined, we quote
\begin{equation}
\sigma^2_A \equiv \frac{1}{F_{AA}},
\end{equation}
which is the forecasted error with all other parameters fixed.
We evaluate the derivatives in Eq.~(\ref{eq:fisher}) at the fiducial $\Lambda$CDM model defined in Sec.~\ref{sec:calcresponse}.  For $A$ we choose the value for which a $2\sigma_A$ detection
is possible which avoids problems with defining the Fisher matrix at $A=0$ discussed below.
For the multipole ranges we choose $L_{\rm min} ={\rm max}(\fsky^{-1/2},2)$, due to the sky cut, and $L_{\rm max}=100$, due to the breakdown of the SU approximation, unless otherwise specified.   

Results for these fiducial choices are given in Table~\ref{tab:2sigma}. Note that for the {CMB}-S4 experiment we find that the curvaton-motivated value of $A\approx16.5$ [realized in the scenario ($b_{\rm by}, c_{\rm before}$)] can in principle be detected at more than
$3\sigma$ statistical significance.   Even for currently available data from WMAP and Planck, the expected limits
are substantially stronger than those determined in Ref.~\cite{Grin:2013uya} for WMAP.   We shall see that in large part these improvements are due to the addition of cross-spectra for correlated CIP modes.

\subsection{Forecast dependencies}
\label{sec:dependencies}

Now let us examine the dependence of these results for the CIP detection threshold
 on various aspects of the data and Fisher matrix assumptions: the maximum and minimum CIP multipole $L_{\rm max}$
and $L_{\rm min}$, the maximum CMB multipole used in the CIP reconstruction
$l_{\rm CMB}$, the impact of including CIP and CMB cross-correlations,  the fiducial
CIP amplitude $A$, and the impact of CMB polarization measurements on reconstruction
and cross-correlation.  In each case, we vary the assumptions one at a time from
the fiducial choices in Sec.~\ref{sec:fisher}.

\begin{figure}
\includegraphics[scale=0.6]{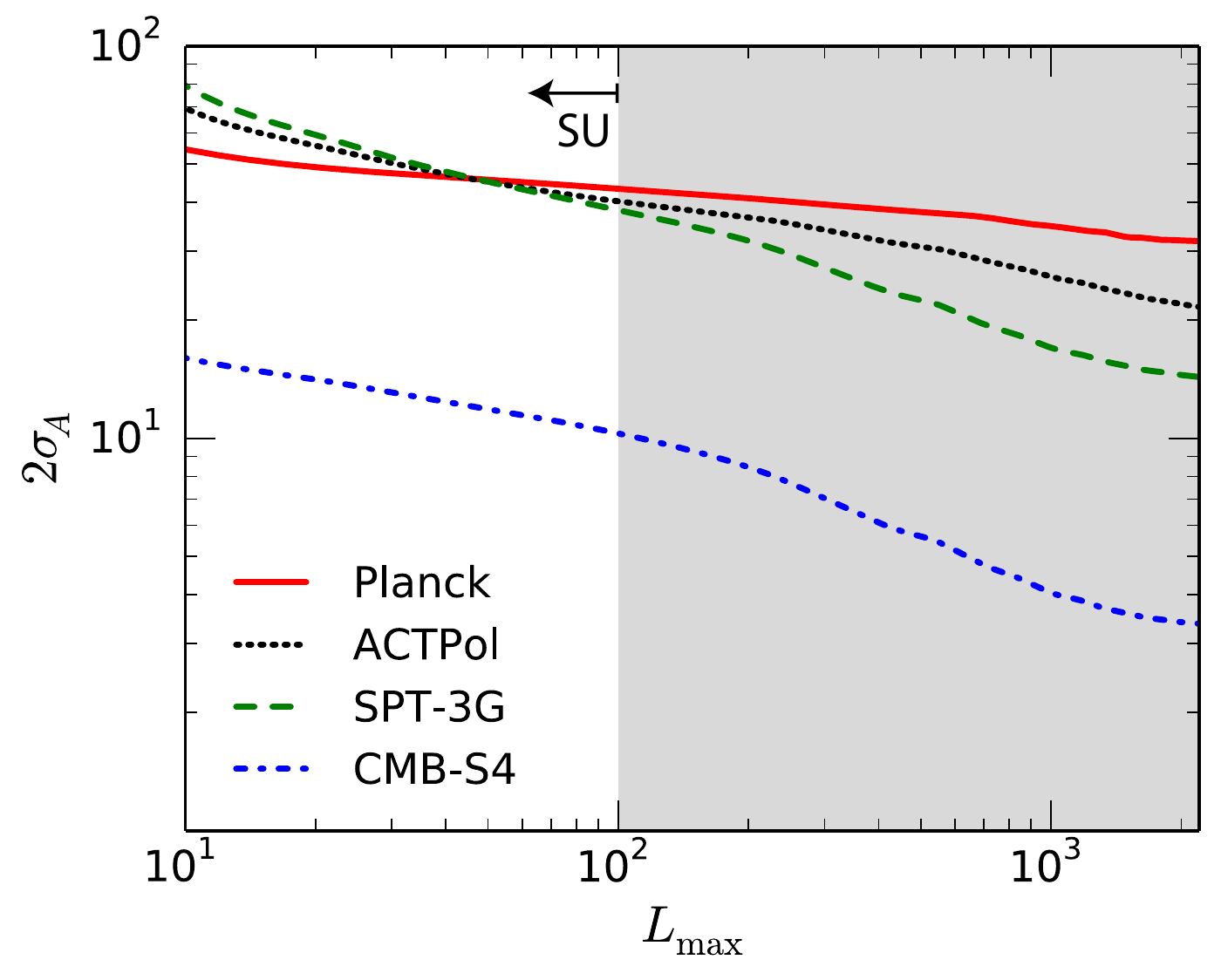}
\caption{\label{fig:lmax} 2$\sigma$  detection threshold 
as a function of the maximum CIP multipole $L_{\rm{max}}$
from all spectra  for Planck, ACTPol, SPT-3G, and CMB-S4 as a function of $L_{\rm{max}}$. The shaded band for $L_{\rm max}>100$ represents the limit of the SU approximation which we take
in all other results.   Other parameters are set to the fiducial choices of Sec.~\ref{sec:fisher}.
 }
\end{figure} 

We begin with the multipole ranges.  Recall that the fiducial CIP maximum multipole  $L_{\rm max}=100$ is chosen to correspond to the angular scale across which sound waves have traveled by the end of the recombination era.  As discussed in Appendix~\ref{sec:wprop}, longer-wavelength CIP modes or smaller
multipoles can be considered as SU variations in cosmological parameters.
Since the breakdown of this approximation occurs within a factor of a few of this
scale (see Fig.~\ref{fig:bias}), we show in Fig.~\ref{fig:lmax}
 the dependence of the $2\sigma$ detection threshold
 on $L_{\rm{max}}$.
 
 For the Planck experiment, whose CIP reconstruction is dominated by $TT$ estimators,
 the dependence on $L_{\rm max}$ is very mild, consistent with the nearly white-noise
 spectrum of estimators that do not involve $B$ modes shown in Fig.~\ref{fig:recon_noise}.
 For future experiments that have good polarization sensitivity, the noise curves of
 $BT$ and $BE$ can cross the others near $L=100$.    Were it not for the breakdown
 of the SU approximation, the implied limits on $A$ would thereafter improve
 substantially, yielding far better constraints with polarization reconstruction and
 with cross-correlation than without.  As discussed below, improvements from cross-correlation depend strongly on the multipole at which the signal is extracted.   With the
 fiducial $L_{\rm max}=100$, polarization reconstruction yields improvements in $N_{L}^{\Delta \Delta}$ of order unity rather than orders of magnitude.
 \begin{figure}
\includegraphics[scale=0.6]{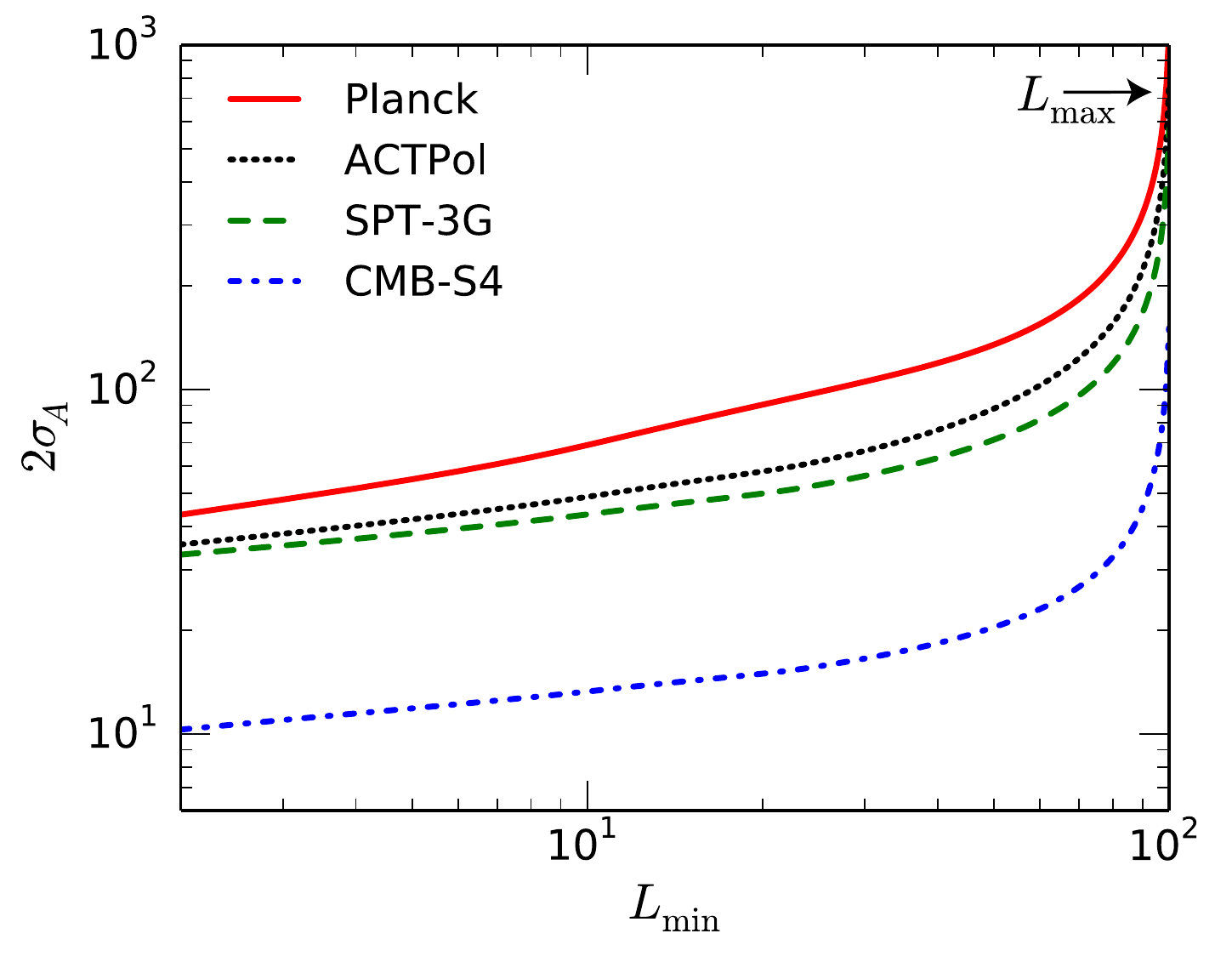}
\caption{\label{fig:lmin} 2$\sigma$ detection threshold  as a function of the minimum CIP multipole $L_{\rm{min}}$ from
all auto- and cross-correlations for Planck, ACTPol, SPT-3G, and CMB-S4.    Other parameters are set to the fiducial choices of Sec.~\ref{sec:fisher}.
}
\end{figure}

Given that non-$B$-mode based CIP reconstruction has the highest signal-to-noise ratio
at the lowest multipoles, it is also interesting to examine the dependence of $2\sigma_A$
on $L_{\rm min}$ with 
$L_{\rm max}=100$
(see Fig.~\ref{fig:lmin}).   In addition to variations due to the 
details of the survey geometry, real experiments
are limited by systematics, $1/f$ noise, and foreground subtraction that can compromise their ability
to probe small multipoles.  As expected, the experiments that are the least dependent
on $B$ modes are the most affected by $L_{\rm min}$.   For Planck, setting
$L_{\rm min}=10$ degrades limits by a factor of 1.6, while for CMB-S4, this choice degrades limits by a factor of $\sim1.3$.

\begin{figure}
\includegraphics[scale=0.6]{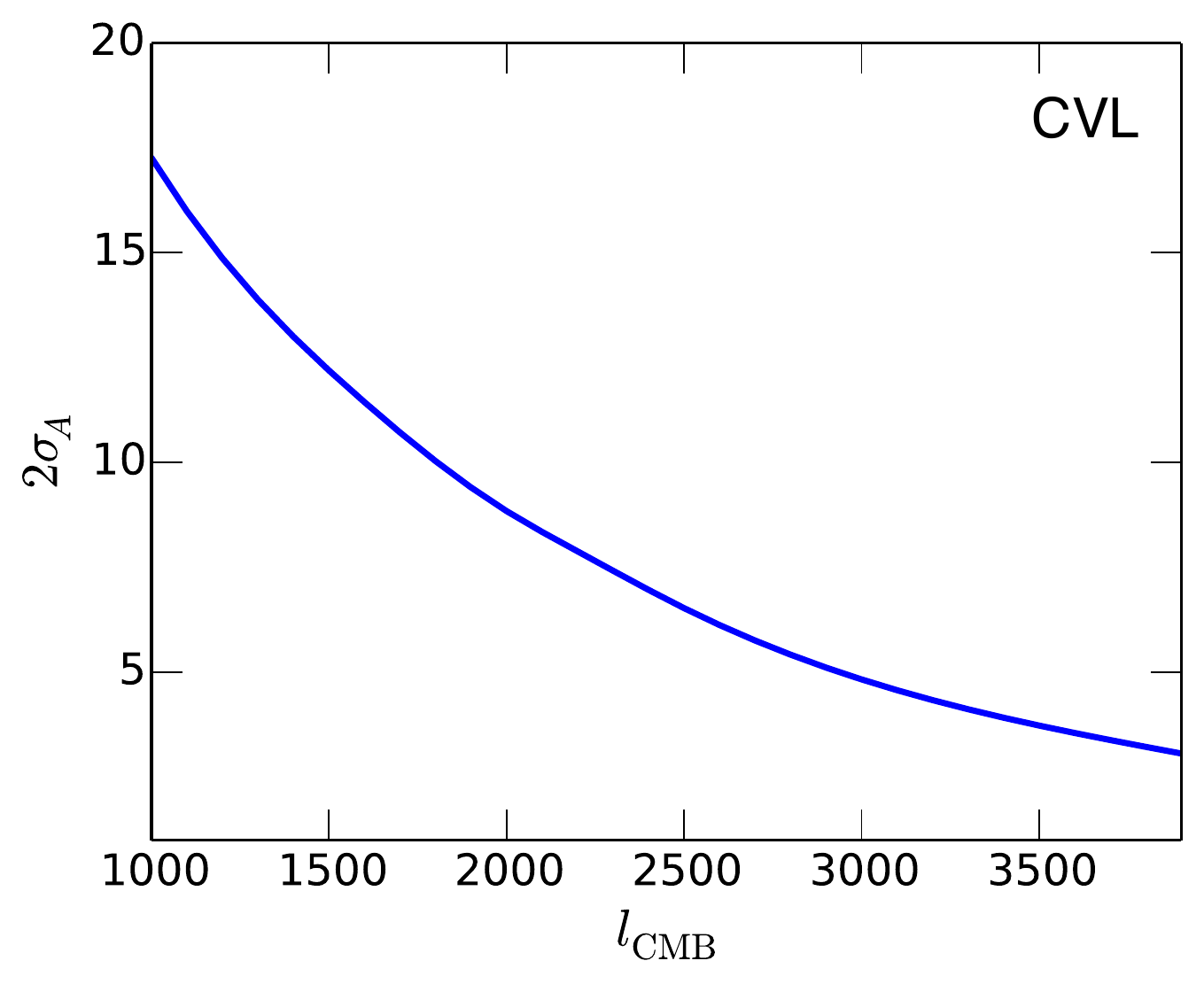} 
\caption{\label{fig:lcmbmax} 2$\sigma$ detection threshold for the CVL case as a function of the maximum CMB multipole $l_{\rm{CMB}}$ used in reconstruction.   Other parameters are set to the fiducial choices of Sec.~\ref{sec:fisher}.  
}
\end{figure}

We have also assumed that CIP reconstruction will be limited to CMB multipoles smaller
than 
$l_{\mathrm{CMB}}=2500$ beyond which the primary anisotropy is severely Silk damped.
For $TT$-based reconstruction, foreground contamination will make information in
higher multipoles difficult to extract regardless of instrument sensitivity.   This is not
necessarily the case for polarization~\cite{Crites:2014nma}, and so in Fig.~\ref{fig:lcmbmax}, we plot the dependence on $l_{\mathrm{CMB}}$ for an ideal CVL 
experiment.  By $l_{\rm CMB}=4000$, the detection threshold
improves to $2\sigma_A\approx 3$.   This level of sensitivity would begin testing the second largest
predicted amplitude of the curvaton dominated scenarios $A=-3$ [realized in the scenario $(b_{\rm before}, c_{\rm by})$] and models with
admixtures of inflaton fluctuations.  Recall that these two largest cases are also the only ones where the effective CMB isocurvature mode can cancel large-angle $TT$ power.

Next we examine the impact of CIP cross-correlation with the CMB and the choice
of the fiducial value of $A$.  
In the  reconstruction noise-dominated regime, cross-correlation helps extract the signal from the noise.
In principle, we can evaluate this improvement by comparing Fisher matrix
errors utilizing only the auto-spectrum ($\alpha = \Delta\Delta$) in Eq.~(\ref{eq:fisher}) with the full result.
There is one important subtlety of Fisher errors that we must address first.
Given that $C_L^{\Delta\Delta} \propto A^2$, it is clear from Eq.~(\ref{eq:fisher})  that 
$F_{AA}\propto A^2$ or $\sigma_A \propto A^{-1}$, which diverges as $A\rightarrow 0$.    On the other hand, the corresponding
 limit on
$A^2$, $\sigma_{A^2}$ remains finite:
\begin{equation}
\sigma_{A^2} = 2 | A | \sigma_A .
\end{equation}
Of course a finite upper limit on $A^2$ implies a
finite limit on $A$ as well, in spite of the Fisher estimate.    The Cramer-Rao bound
only guarantees that the Fisher estimate
gives the best possible errors of an  {\it unbiased} estimator.  When the data only
provide an upper limit, such an estimator can be substantially suboptimal.    For this reason,
we choose for our fiducial signal the point where $A=2\sigma_A$, which represents 
a low significance detection rather than an upper limit.  

\begin{figure} 
\includegraphics[scale=0.6]{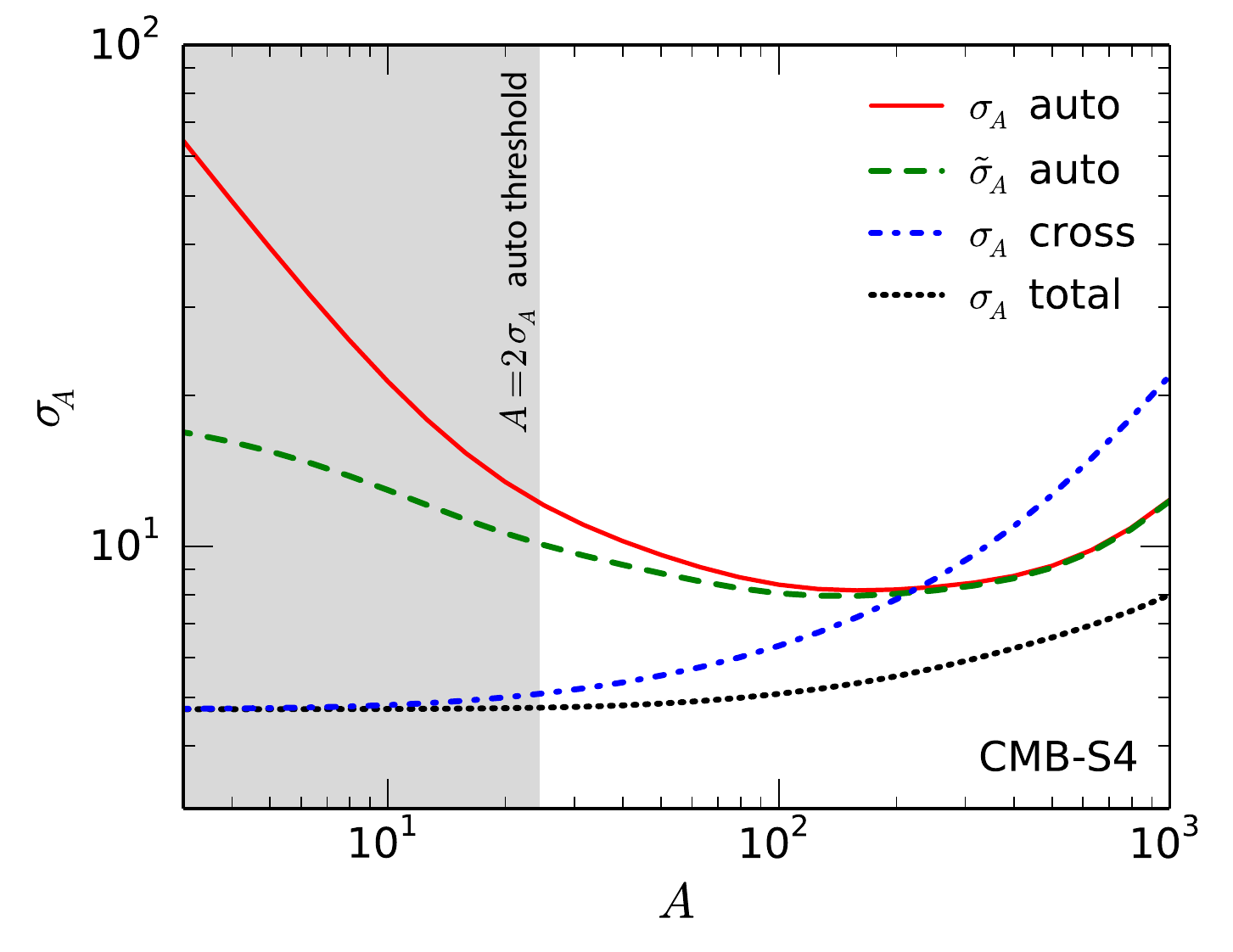}
\caption{\label{fig:sigma_vs_A} Fisher error $\sigma_{A}$  vs $A$ for CMB-S4 from combinations of auto- ($\Delta\Delta$) and cross-spectra ($T\Delta$, $E\Delta$).  Also shown is $\tilde\sigma_A$, the Fisher error implied from $\sigma_{A^2}$ for auto-spectra from Eq.~(\ref{eq:tilde_sigma}).  The two auto-spectra analyses agree well for $A> 2\sigma_A$, which
defines the regime where we can meaningfully compare various Fisher results.  Cross-spectra increasingly dominate the total at low signal-to-noise ratio $A/\sigma_A$ but also improve the total result at high signal-to-noise ratio by reducing sample variance.  All other parameters 
are set according to Sec.~\ref{sec:fisher}.}
\end{figure}

To ensure that our Fisher estimates of $\sigma_A$ from auto-correlation are not misleading,
consider an alternate definition of the error on $A$ that corresponds to a mapping of
the upper limit on $A^2$:
\begin{eqnarray} \label{eq:tilde_sigma}
	\tilde\sigma_A &\equiv & { \sqrt{A^2+\sigma_{A^2}}-|A|} \nonumber\\
	&=& { \sqrt{A^2+2 |A| \sigma_A}-|A|}.
\end{eqnarray}
This quantity remains finite at $A\rightarrow 0$ as expected.
In Fig.~\ref{fig:sigma_vs_A}, we compare  $\sigma_A$ and $\tilde{\sigma}_A$ of 
the $\Delta\Delta$ auto-spectrum
for the {CMB-S4} experiment.   The two estimates agree well as long as $|A| \gtrsim 2\sigma_A(A)$ (unshaded region).   
This is the reason we quote our primary results as the value of $A$ at the detection threshold
$A=2\sigma_A$.

We also show here the results obtained if \textit{only} cross-spectra ($T\Delta$ and $E\Delta$) are used, for comparison with the total. In the $A\rightarrow 0$ limit, they dominate the Fisher information leading $\sigma_{A}$ to be finite and independent of the fiducial $A$ value
in this limit.

Also interesting is
the signal-dominated large-$A$ regime where $C_L^{\Delta\Delta} \gg N_L^{\Delta\Delta}$.  
Although the cross-spectra alone provide worse limits than the auto-spectrum alone, 
the total is better than what one would expect by summing their independent information content.   This is because having the auto- and cross- correlation helps eliminate the
sample variance of the Gaussian random curvature fluctuations $\zeta$.  Indeed,
if the CIP and CMB modes were perfectly correlated, sample variance could be eliminated
entirely.

For the S4 experiment, Fig.~\ref{fig:sigma_vs_A} shows that the cross-spectra
improve the detection threshold for $A$ by a factor of 2.3. It is interesting
to trace this improvement back to the level of correlation between the CIP and CMB 
modes.   By assuming the noise-dominated regime $C_L^{\Delta\Delta} \ll N_L^{\Delta\Delta}$, as appropriate for a
first detection, we can approximate the auto-spectra errors as 
\beq \label{eq:sigma_DD} 
	\sigma^{-2}_A \Big|_{\Delta\Delta} \approx \sum_L   \frac{2L+1}{2} f_{\rm sky}  \left(\frac{2}{A}  \frac{C_L^{\Delta\Delta}}{N_L^{\Delta\Delta}} \right)^2
\eeq
and compare this to the cross-spectrum where $X \in T,E$:
\beq \label{eq:sigma_XD} 
	\sigma^{-2}_A \Big|_{X\Delta} \approx \sum_L  ( {2L+1} ) f_{\rm sky} \frac{1}{A^2} \frac{C_L^{\Delta\Delta}}{ N_L^{\Delta\Delta}}\frac{C_L^{XX}}{\tilde C_L^{XX}} (R_L^{X\Delta})^2, 
\eeq
where we recall that $R_L^{X\Delta}$ is the cross-correlation coefficient shown in Fig.~\ref{fig:R}.
As expected, cross-correlation is more important when the reconstruction signal-to-noise ratio
$C_L^{\Delta\Delta}/N_L^{\Delta\Delta}$ is smaller.

For a detection threshold $A = 2\sigma_A$, we can estimate this ratio and hence how the improvement scales with experimental assumptions.
For the auto-spectrum this {threshold occurs} when
\beq
	\frac{C\DD}{N\DD} \sim \frac{1}{ L_{\Delta} f_{\rm sky}^{1/2}}, 
\eeq
where $L_\Delta$ is a representative multipole, {roughly the $L$ value by which the
Fisher sum accumulates half its total value}.   For this level of
signal, the cross-spectra would give better constraints by a factor of 
\beq \label{eq:improvement_cross_auto}
	\frac{\sigma^{-2}_A |_{X\Delta} }{\sigma^{-2}_A |_{\Delta\Delta} }
\sim  L_{\Delta} f_{\rm sky}^{1/2}\frac{C_L^{XX}}{\tilde C_L^{XX}}   (R_L^{X\Delta})^2.
\eeq
Given that the correlation coefficient averaged over a sufficiently large range in $L$
is always of order unity, we can now see that the improvement due to adding the
cross-correlation depends on very few aspects of the experiment.
  For an experiment
whose CIP reconstruction is dominated by CVL $TT$ measurements like Planck, the typical CIP multipole in the signal is $L \lesssim 10$ and the improvement is limited (1.9 for Planck). 
For an experiment with good polarization sensitivity and sky coverage, the improvement
can be larger due to both the higher $L$ out to which the signal can be detected and
the addition of the $E\Delta$ cross-spectrum. In fact, improvements are ultimately limited
by the SU approximation $L \lesssim 100$.  For the S4 experiment, the improvement
of a factor of 2.3 from the cross-spectra comes partially from $E\Delta$.  Without $E\Delta$,
the 2$\sigma_A$ threshold goes from 10 to 15, and hence  polarization plays a significant role
in making a $3\sigma$ detection of $A \approx 16.5$ possible.

Finally, it is interesting to compare these forecasted constraints with existing CMB constraints to CIPs, which come from $C_{L}^{\Delta\Delta}$ alone and do not apply information from cross-correlations of $T/E$ with $\Delta.$
The latest WMAP 95\% C.L. upper limit on a scale-invariant spectrum of CIPs is $L(L+1)C\DD \leq 0.011$ \cite{Grin:2013uya}.  For correlated CIP modes, this corresponds to an upper limit of $A^2 \leq (808)^2$. 
The Fisher forecast from the auto-spectrum predicts that the $2\sigma$ detection threshold in $A^2$ is $A^2 = 2\sigma_{A^2} 
= (469)^2$. 
Note that in the Fisher approximation a 2$\sigma$ detection in $\sigma_{A^2}=2 A^2$ implies
$\sigma_A = 4 A$ (for example, the 2$\sigma$ threshold in $A$ for WMAP is $A = 264$). 
The Fisher value in this case underestimates the actual errors by a factor of $3.0$ in $A^2$ or $1.7$ in $A$, which should be borne in mind when considering forecasted errors.  We have isolated the root of this discrepancy to the difference between the forecast instrument noise obtained from Eq.~(\ref{eq:instrument_noise}) and the true WMAP instrument noise at map level.

The Fisher auto-result can also be compared with the signal-to-noise forecast in Ref.~\cite{Grin:2011tf} for a scale-invariant CIP in the WMAP 7-year data release. There, to be maximally conservative, instrument noise was computed assuming a single differencing assembly.  The resulting forecast, $S/N = 300 \Delta_{\rm cl}^2$,
corresponds to $\sigma_{A^2} = 1.8\times10^5$ for the correlated CIP in the $N\DD \gg C\DD$ regime. Our Fisher forecast $\sigma_{A^2}=~4.2\times10^4$ is a factor of 4.3 lower, mostly due to lower values of $N\DD$ calculated for the full multiple-differencing-assembly, 9-year experiment.

\section{Conclusions}
\label{sec:conclude}
In the curvaton model, quantum fluctuations of a spectator field during inflation  seed the primordial curvature perturbation $\zeta$ after inflation and in the process can produce correlated isocurvature fluctuations from its decay. In some curvaton-decay scenarios, the usual adiabatic and total matter isocurvature perturbations are accompanied by relatively unconstrained compensated isocurvature perturbations (CIPs) between baryons and dark matter. In the curvaton model, CIPs are correlated with the adiabatic fluctuations with amplitude given by $S_{b\gamma}^{\rm CIP}=A\zeta$.  

The most interesting (and observationally allowed) scenarios are those where baryon number is generated by curvaton decay, while cold dark matter is generated before; or where baryon number is generated before curvaton decay, and cold dark matter is generated directly by curvaton decay. These cases yield CIP amplitudes of $A\approx 16.5$ and $A = -3$, respectively.  By modulating the propagation of acoustic waves during the tightly coupled epoch, CIPs induce detectable off-diagonal two-point correlations in the CMB \cite{Grin:2011tf}.  The correlation with the dominant adiabatic mode means that the cross-power spectrum between the CIP estimators and the CMB fields
themselves can in principle be used to probe very small values of $A$ where the auto-correlation is too noisy for detection.

	In this work, we obtain the expected amplitude of fully correlated CIPs in the different curvaton-decay scenarios relevant to dark matter and baryon number production. The sensitivity to $A$ of seven different CMB experiments and the ideal cosmic-variance-limited (CVL) case was computed using the Fisher information and applying a more refined calculation of CIP reconstruction noise than past work \cite{Grin:2011tf}. We find that the validity of the implicit separate-universe approximation made in previous work \cite{Grin:2011tf} requires a cut $L\lesssim100$ on the multipole index of the reconstructed CIP multipole moments $\hat{\Delta}_{LM}$. While this cut does not affect existing limits to CIPs like Ref.~\cite{Grin:2013uya}, it is important for predictions of future sensitivity, particularly for precise future CMB polarization experiments.
	
Large-scale CMB temperature anisotropies are correlated with the large-scale primordial curvature perturbation $\zeta$, and so cross-correlating the reconstructed CIP with CMB temperature maps can improve the detection threshold for $A$ by a factor of $1.7$-$2.7$ depending on the experiment.  The smallest values in this range apply for $TT$-dominated experiments such as WMAP or Planck, and we expect that the upper limits to $A$ from Ref.~\cite{Grin:2013uya} would improve by a factor of $\simeq 1.7$ if those CIP maps were cross-correlated with large-scale temperature maps.  For a CVL experiment out to multipoles $l<2500$, the improvement by a factor $\sim 2.7$ is largely independent of the instrument details and most sensitive to $L_{\Delta}$, the multipole below which the majority of the S/N comes from. Since the $TT$ estimator noise has a steeper slope in $L$ than the $BT$ estimator, polarization-dominated experiments will naturally have S/N up to a larger $L_{\Delta}$.

The planned CMB-S4 experiment will approach the cosmic-variance limit for polarization. As a result, it could detect the $A\approx16.5$ scenario (the largest value attainable in curvaton CIP scenarios) with more than $3\sigma$ significance.   If polarized foregrounds are negligible or can be removed so that CIP reconstruction can be performed with $l_{\rm CMB}\sim4000$~\cite{Crites:2014nma}, the sensitivity to $A$ of a CVL experiment will dramatically improve.  In the cosmic-variance limit, this would allow the $A\approx16.5$ scenario to be detected with $\sim 11\sigma$ significance, and possibly test the second largest CIP scenario $|A| = 3$, as well as models with admixtures of inflaton fluctuations.
	
A detection of fully correlated CIPs could discriminate between the different curvaton-decay scenarios. The largest correlation $A \approx16.5$ arises in the $(b_{\rm by}, c_{\rm before})$ scenario, where the baryon number is created \textit{by} the curvaton decay and the CDM number \textit{before} the decay. This is the last observationally permitted scenario in which dark matter is produced before curvaton decay, as the other cases are already ruled out by the matter isocurvature constraints. A detection of $A\approx16.5$ would provide strong support  for the $(b_{\rm by},c_{\rm before})$ curvaton scenario, in which the dark matter must be produced before curvaton decay, pointing us towards novel dark-matter production mechanisms prior to curvaton decay.  This would also hint that baryon number generation is connected to the physics of a spectator field during inflation. This case would also
predict a level of local non-Gaussianity of $f_{\rm nl}\approx 6$~\cite{Sasaki:2006kq}
that might be used to confirm a measurement from CIPs.  Indeed, the Planck temperature   ($f_{\rm nl} = 2.5\pm5.7$, 68\% C.L.)
and preliminary polarization $(f_{\rm nl} = 0.8\pm5.0$, 68\% C.L.) constraints are already close
to this predicted amplitude \cite{Ade:2015ava}.
If $A\approx 16.5$  is ruled out by either means, we would know that in the curvaton model, dark matter is either directly produced by curvaton decay or (thermally, from the relativistic plasma) after curvaton decay.   

Challenges remain for future work, in particular, a precise evaluation of biases in correlated CIP measurements from off-diagonal correlations induced by weak gravitational lensing, and the generalization of the expressions here to models where inflaton and curvaton contributions to $\zeta$ are more comparable. Our work may pave the way for future CMB measurements to uncover the physics of curvaton decay.

\begin{acknowledgments}
C.H. and W.H. were supported by the Kavli Institute for Cosmological Physics at the University of Chicago through grants NSF PHY-1125897 and an endowment from the Kavli Foundation and its founder Fred Kavli, by U.S.~Dept.\ of Energy contract DE-FG02-13ER41958 and NASA ATP NNX15AK22G. DG is funded at the University of Chicago by a National Science Foundation Astronomy and Astrophysics Postdoctoral Fellowship under Award NO. AST-1302856. We thank B.~Benson, T.~Crawford, A.~L.~Erickcek, C.~Gordon, M.~Kamionkowski, M.~LoVerde, A.~Manzotti,  S.~Meyer, J.~Sievers, K.~Sigurdson, and T.~L.~Smith for useful discussions. We are especially grateful to A.~L.~Erickcek and T.~L.~Smith for a thorough reading of the manuscript.\end{acknowledgments}

\appendix
	
\section{Wave Propagation in an Inhomogeneous Medium }
\label{sec:wprop}
Adiabatic acoustic waves in the CMB propagate on a background that is spatially modulated
by the presence of the CIP mode $S_{b\gamma}^{\mathrm{CIP}}(\bx)$.  Here we present a 
simplified model of this system to test the domain of validity of the separate-universe (SU) approximation
introduced in Ref.~\cite{Grin:2011tf}.   

In this simple model, we consider an acoustic wave in fractional temperature fluctuations $T$ propagating
in a medium with temporally constant, but spatially modulated sound speed
\begin{equation}
\ddot T - c_s^2 \left[1 + \frac{d \ln c_s^2}{d \Delta}\Delta({\bx}) \right]\nabla^2 T = 0.
\label{eqn:wave}
\end{equation}
Note that the qualitative difference between this model and the SU model 
used in Eq.~(\ref{eqn:Clderiv}) is that the Taylor approximation {for $c_{s}^{2}(\mathbf{x})$ is employed for the medium itself, rather than for the observables after the acoustic mode has propagated through the medium.}  Since the medium is only weakly inhomogeneous,
we can solve the wave equation by iteration. By expanding
\begin{equation} 
T = T_0 + T_1 + \cdots,
\end{equation}
Eq.~(\ref{eqn:wave}) becomes 
	\begin{align}
	\ddot T_0 - c_s^2 \nabla^2 T_0 &= 0, \nonumber\\
		\ddot T_1 - c_s^2 \nabla^2T_1 &= c_s^2 \frac{d \ln c_s^2}{d \Delta} \Delta(\bx) \nabla^2 T_0,
	\end{align}
where overdots are derivatives with respect to the time variable $\eta$.    
For the zeroth-order solution, we take
$T_0 = -\zeta/5$, $\dot T_0=0$ as the initial condition and solve for the Fourier modes
\begin{equation}
T_0(\bk,s) =- \frac{1}{5} \zeta(\bk) \cos(k s),
\end{equation}
where the sound horizon is 
\begin{equation}
s(\eta) =  c_s \eta .
\end{equation}

Given the solutions to the homogeneous equation, the solution for $T_1$ is
	\begin{align}
	T_1(\bk, s) = & \frac{1}{5} \frac{d \ln c_s^2}{d \Delta} \int \frac{d^3 k'}{(2\pi)^3}  
	\Delta(\bk-\bk') \zeta(\bk')
\\	& \times \frac{k'^2}{k^2 - k'^2} \left[
	\cos(k' s) - \cos(k  s)\right].		\notag
	\end{align}
Note that $T_1$ responds as an oscillator subject to an external force given by the modulation and the unmodulated solution.   In particular
when $k'=k$ the oscillator is driven at its natural frequency leading to an enhanced response.

The two-point correlations with fixed modulating mode can likewise be expanded to first order in the modulation
\begin{align}
\langle T(\bk)T( \bk') \rangle_T \approx & \langle T_0(\bk)T_0( \bk') \rangle_T +
R(k,k') \Delta({\bK}),
\label{eqn:toyoffdiag}
\end{align}
where $\bK = \bk + \bk'$. The unmodulated piece is given by
\begin{equation}
\langle T_0(\bk)T_0( \bk') \rangle_T = (2\pi)^3 \delta(\bk + \bk')\cos^2(k s) \frac{P_{\zeta\zeta}(k)}{25},
\end{equation}
and the modulation response function is given by
	\begin{align}\label{eq:R}
	R(k,k') =&  -  \frac{d \ln c_s^2}{d \Delta}\frac{ P_{\zeta\zeta}(k')}{25}  \frac{k'^2}{k^2 - k'^2} \cos(k's)  
\\	& \times \left [\cos(k' s) - \cos(k  s)\right] + {\rm perm.},\notag
	\end{align}
where the permutation refers to $k \leftrightarrow k'$. 
This response should be compared with the SU approximation, where
the Taylor expansion is performed on the solution rather than the medium
\begin{equation}
T^{\rm SU}(\bx,s) =T_0(\bx,s) + \frac{d T_0}{d \Delta}(\bx,s) \Delta(\bx),
\end{equation}
and gives off-diagonal correlations of the form of Eq.~(\ref{eqn:toyoffdiag}) but
with the SU response function
\begin{eqnarray}
 R^{\rm SU}(k,k') &= &\frac{P_{\zeta\zeta}(k)}{25} \cos(k s)\frac{d\cos (ks) }{d \Delta}  
 + {\rm perm.}\\
 \label{eqn:Rsu}
 &=&- \frac{d \ln c_s^2}{d \Delta} \frac{P_{\zeta\zeta}(k)}{25} \frac{ks}{4}  \sin(2ks) +  {\rm perm.}\nonumber
\end{eqnarray} 
Note that  $-\cos(k s)/5$ plays the role of the transfer function.

A comparison shows that the two are only equal in the limit $k' \rightarrow k$. 
To keep track of the differences, let us define
 	\beq
	\epsilon \equiv (k'-k)s
	\eeq
and $x= k s$.	
We can rewrite the response [Eq.~(\ref{eq:R})] as
	\begin{align}\label{eq:Reps}
	R(k,k') =&   \frac{d \ln c_s^2}{d \Delta}\frac{ P_{\zeta\zeta}(k')}{25}  \frac{(x+\epsilon)^2}{x^2 - (x+\epsilon)^2} \sin(\epsilon/2)  
\\	& \times[\sin(2x +3\epsilon/2) - \sin(\epsilon/2)]+ {\rm perm}.\notag
	\end{align}
It is clear that the validity of the SU approximation  requires $|\epsilon|\ll 1$.   It is not sufficient for {the} wavelength of the acoustic mode to be much smaller than the
modulating mode ($k/K \gg 1$).   The criterion for a coherent driving of the oscillator is that the phase error introduced by $\epsilon$ itself be small.

Now, let us consider the relevant case for reconstruction where $k \gg K$ or $x \gg \epsilon$ and there are many pairs of acoustic modes (satisfying $|\mathbf{k}|\simeq |\mathbf{k}'|\simeq k$) that
can be used to measure the modulation.
If $P_{\zeta\zeta}$ is a featureless power law, the response can be simplified considerably, yielding
	\begin{equation}\label{eq:Rlim}
	R(k,k') \approx 	 - \frac{d \ln c_s^2}{d \Delta} \frac{P_{\zeta\zeta}(k)}{25} \frac{ks}{2}  \frac{\sin\epsilon}{\epsilon} {\sin(2ks+\epsilon) } , 
	\end{equation}
which leads to both a damping of the response and a decoherence in the phase for $|\epsilon| > 1$.  

Next, consider the impact of the reduced response on the estimator of the modulation mode.   
In the SU approximation, we can use Eq.~(\ref{eqn:toyoffdiag}) to obtain the minimum-variance CIP estimator:\begin{eqnarray}
\hat \Delta({\bK})&=&N_{K}^{\rm SU}\int \frac{ d^{3} k}{(2\pi)^3} \frac{T({\bk})T({\bk'})R^{\rm SU}(k,k')}{\tilde P_{TT}(k)\tilde P_{TT}(k')},\nonumber\\
\left(N_{K}^{\rm SU}\right)^{-1}&=& \int  \frac{ d^{3} k}{(2\pi)^3}  \frac{\left[R^{\rm SU}(k,k')\right]^{2}}{\tilde P_{TT}(k)\tilde P_{TT}(k')}, \label{eq:toy_estimator}
\end{eqnarray}
where $\bk + \bk' = \bK$ and
\begin{equation}
\tilde P_{TT}(k) =\cos^2(ks) \frac{P_{\zeta\zeta}(k)}{25} + N_{TT}(k),
\end{equation}
with $N_{TT}$ as the  noise power spectrum from measurement errors.    Beyond the
SU approximation, this estimator is biased: 
\begin{eqnarray}
b_{K}(Kr_{s})&\equiv &\frac{ \langle \hat{\Delta}{(\bK)} \rangle_T}{\Delta(\bK)}\nonumber\\&=&N_{K}^{\rm SU}\int \frac{ d^{3} k}{(2\pi)^3} \frac{R(k,k')R^{\rm SU}(k,k')}{\tilde P_{TT}(k)\tilde P_{TT}(k')}.\label{eq:bexact}
\end{eqnarray}

To estimate this bias, we can first determine $\epsilon$ for
each pair of modes that
satisfies $\bk + \bk' = \bK$.     Defining $\bk \cdot {\bK} = \mu k K$
and assuming $k \gg K$, 
we have
\begin{equation}
\epsilon \approx - K s \mu.
\end{equation}
Note that $\epsilon$ does not depend on $k$, but only on the angle of $\bk$ with $\bK$. 

We can approximate the minimum 
variance estimator in Eq.~(\ref{eq:toy_estimator}) by ignoring variations in the weights due to the $k$ dependence of $\tilde P_{TT}(k)$, in
particular due to the unphysical zeros in  power which would be filled in by the Doppler
effect and projections in a real observable.
Given the difference of the true- and separate-universe responses, this estimator
would be biased as 
\begin{eqnarray}
b_K &\approx& \frac{ \langle  \int_{-1}^{1} d\mu  
\frac{\sin \epsilon}{\epsilon} {\sin (2x + \epsilon)}[\sin(2 x) + \sin(2 x + 2\epsilon)]
	\rangle_x} 
	{ \langle  \frac{1}{2} \int_{-1}^{1} d\mu [\sin(2 x) + \sin(2 x + 2\epsilon)]^2
	\rangle_x }
\nonumber\\
&=& \frac{2 {\rm Si}( 2 K s) }{2 Ks + \sin(2 K s)},\label{eq:bk}
\end{eqnarray}
where $\langle ...\rangle_x$ denotes an average of a cycle of the oscillation in $x$.

\begin{figure}[t]
\includegraphics[width=3.4 in]{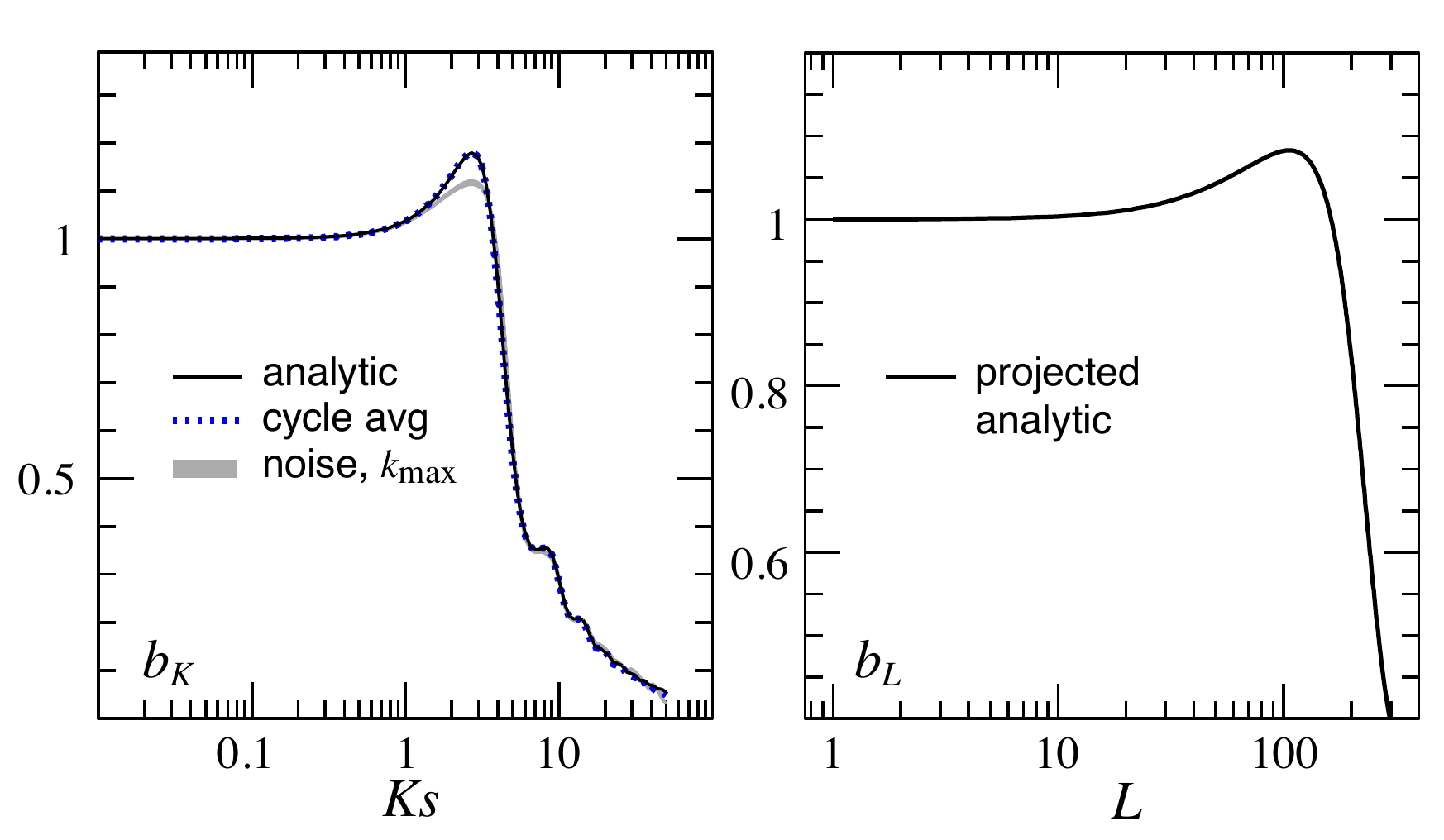}
\caption{Reconstruction bias in $K$ (left) and in $L$ (right) space.  Left: Analytic expression of Eq.~(\ref{eq:bk}) is in excellent agreement with the evaluation of Eq.~(\ref{eq:bexact}) using the same
cycle-averaged replacement assumption 
$\tilde P_{TT}(k)\to \langle \tilde P_{TT}(k) \rangle_{x}$.   The gray range shows the full calculation for a wide family of Poisson noise power spectra and 
$k_{\rm max}$.   Right: Projected bias using the analytic expression from 
Eq.~(\ref{eq:toy2dbias}) with $\theta_*=0.01$, showing that $L<100$ is nearly unbiased.}
\label{fig:bias}
\end{figure}

Near $K s \approx 1$, the estimator becomes slightly positively biased, $b_K>1$,
but quickly falls to $b_K \ll 1$ for larger values. Thus, the estimator essentially low-pass 
filters the modulation field, allowing through modes that are larger than the sound horizon. We show the behavior of $b_K$ in Fig.\ \ref{fig:bias} (left), calculated using both Eq.~(\ref{eq:bexact}) (shown as a range for different noise models, and variety of $k_{\rm max}/K \gg 1$) and the analytic approximation Eq.~(\ref{eq:bk}).   The former is nearly independent of assumptions
and agrees very well with the latter except for a small range around $Ks \sim 3$.
Also shown is the result of evaluating Eq.~(\ref{eq:bexact}), with $N_{TT}=0$ for definiteness, but with the same cycle-averaged assumption  $\tilde P_{TT}(k)\to \langle \tilde P_{TT}(k) \rangle_{x}$  that was made in the analytic approximation.   The agreement with the analytic approximation 
shows that the overshoot around $Ks \sim 3$ is due to this average.
 With this average, the unphysical zero crossings of the sound wave are eliminated in both Eqs.~(\ref{eq:bexact}) and (\ref{eq:bk}), and in that sense is closer to a physical model than the full calculation of
Eq.~(\ref{eq:bexact}).   In any case, the analytic model captures the main feature
of the bias which is a sharp cutoff in sensitivity for $Ks \gg 1$.

For angular projections of the acoustic waves and the modulation mode, this reduced response and the corresponding bias is somewhat larger. In the flat-sky approximation, the angular modulation mode
is related to the spatial one at an epoch $s$ as 
\begin{equation}
\Delta(\bL) = \frac{1}{D^2} \int \frac{ d K_\parallel}{2\pi}\Delta(\bK_\perp= \bL/D, K_\parallel)
e^{i k_\parallel D},
\end{equation}
where $D$ is the distance to the observed surface, and $\parallel$ and $\perp$ denote directions with respect to the line of sight.  Defining $T(\bl)$ analogously,
we again have off-diagonal correlations of angular moments
\begin{eqnarray}
\langle T(\bl ) T(\bl') \rangle  &=& \langle T_0(\bl ) T_0(\bl') \rangle_T
+\frac{1}{D^4} \int \frac{ d k_\parallel}{2\pi} \int \frac{ d k_\parallel'}{2\pi} e^{i K_\parallel D}
\nonumber\\
&&\times R(k,k') \Delta(\bL/D,K_\parallel),
\label{eqn:angular}
\end{eqnarray}
where $\bL  =\bl + \bl'$ and 
\begin{eqnarray}
\langle T_0(\bl) T_0( \bl') \rangle_T &= &(2\pi)^2 \delta(\bl +  \bl')
\int \dfrac{ dk_\parallel}{2\pi} \cos^2 (k s) \nonumber\\
&&\times \dfrac{ P_{\zeta\zeta}(l/D,k_\parallel)}{25 D^2}.
\label{eqn:T0angular}
\end{eqnarray}
Note that here we take the angular observable as the value of the
local temperature field on the surface.   For the real case of CMB temperature
anisotropy, the Doppler effect also contributes and suffers even greater 
 projection effects.

  In the separate-universe approximation for $R$ in Eq.~(\ref{eqn:Rsu}), Eq.~(\ref{eqn:angular}) reduces to
\begin{equation}
\langle T(\bl ) T(\bl') \rangle_T  = \langle T_0(\bl ) T_0(\bl') \rangle_T  + R^{\rm SU}(l,l') \Delta(\bL).
\label{eqn:angularsu}
\end{equation}
The angular response function
\begin{eqnarray} 
R^{\rm SU} (l,l') &=& - \frac{d \ln c_s^2}{d \Delta} \int \frac{d k_\parallel}{2\pi}\frac{ks}{4}  \sin(2ks)  \frac{P_{\zeta\zeta}(l/D,k_\parallel)}{25 D^2}
\nonumber\\
&&+  {\rm perm.}\
 \end{eqnarray}
 is again the derivative of the transfer function implied by Eq.~(\ref{eqn:T0angular}).   Likewise, the SU approximation holds for triangles in the integrals where $|\epsilon | \ll 1$. Due to the projection, however,
even if the projected 2D triangles in the transverse plane satisfy the analogous criteria, the 3D triangles that compose the estimators may not, since $K \ge K_\perp$.

In the exact expression of Eq.~(\ref{eqn:angular}),
the off-diagonal angular multipole pairs are no longer simply proportional to the projected modulation mode $\Delta(\bL)$.   Thus, it is not possible to evaluate the bias in the reconstruction itself. Instead, the bias appears in the auto- and cross-correlations of the reconstructed mode. Since the expressions are cumbersome, we instead estimate projection effects by calculating the bias that would result if the auto-correlation were constructed by the projection of biased estimators of $\Delta(\bK)$.
 
 For a nearly scale-invariant spectrum {($K^3 P_{\Delta\Delta} \approx $ const.)},
 \begin{eqnarray}
C_{L}^{\Delta\Delta} &=& \frac{1}{D^2}\int \frac{d K_\parallel}{2\pi} P^{\Delta\Delta}(
L/D,K_\parallel) \nonumber\\
&\approx&   \frac{2\pi}{D^2} \frac{ K^3 P^{\Delta\Delta}
 }{2\pi^2} \int_{K_{\perp}}^{\infty} \frac {d K}{K^2 \sqrt{K^2 - K_\perp^2}} 
\nonumber\\
 &=& \frac{2\pi}{L^2 } \frac{ K^3 P^{\Delta\Delta}
 }{2\pi^2} .
\end{eqnarray}
Thus, the angular power spectrum gets contributions from modes with $y=K/K_\perp>1$ 
as
\begin{equation}
\frac{d \ln C_L^{\Delta\Delta}}{d y} = p(y) \approx \frac{1}{y^2 \sqrt{y^2-1}}.
\end{equation}

We therefore estimate projection effects by weighting $b_K^2$ accordingly:
 \begin{equation}
b_L^{2}(L\theta_*)= \int_1^\infty  dy\,  p(y) b_K^2( Ks= y L \theta_*) .\label{eq:toy2dbias}
\end{equation} 
Here $\theta_* = s/D$ is the projected acoustic scale. In the $\Lambda$CDM cosmology, $\theta_* \approx 0.01$, and so $L\theta_*=1$ for $L=100$. The resulting projected bias is shown in Fig.~\ref{fig:bias} (right). Clearly the bump near $L\theta_{*}=1$ arises from the bump in the three-dimensional bias plot, Fig.~ \ref{fig:bias} (left). For this toy model of CMB acoustic waves, we see that $L=100$ is a scale at which to safely truncate all estimators of $\Delta$. Since this is only a toy model, we expect that this estimate is only accurate to order unity 
corrections and explore sensitivity to variations in $L_{\rm max}$ in the main text.

\section{Improved reconstruction noise curves}
\label{sec:correct_noise}
The CIP reconstruction methods introduced in Ref.~\cite{Grin:2011tf} are valid as long as $L\lesssim 100$, as discussed in Appendix \ref{sec:wprop}. Aside from changes to precise instrumental noise properties and best-fit cosmological parameters, the curves shown for $N_{L}^{TT}$ and $N_{L}^{EE}$ in Ref.~\cite{Grin:2011tf} are thus valid for all $L\lesssim 100$. We find, however, that numerical errors were made in Ref.~\cite{Grin:2011tf}, {affecting the shape of $N{\DD}$ at scales $L\gsim 100$; the limit to CIP reconstruction ($L\lesssim 100$) imposed by the SU approximation, however, means that these errors are of no practical significance, and have no bearing on the validity of existing CMB limits to CIPs~\cite{Grin:2013uya}.}

Due to an erroneous index in the code employed there, however, a swap took place between the indices $l$ and $l'$ when evaluating Eqs.~(\ref{eq:single_noise}) and (\ref{eq:single_norm}) for $N_{L}^{TB}$ and $N_{L}^{EB}$. For the last two experimental cases considered there [the nearly cosmic-variance-limited (CVL) EPIC mission concept and the actual cosmic-variance limit], the analytic damping envelope reionization prescription of Ref. \cite{Hu:1996mn}, was employed, rather than the constant-$\tau$ prescription employed elsewhere in Ref.~\cite{Grin:2011tf}. For $L\lesssim 100$, we find that together these two errors lead to errors of $\Delta N_{L}^{XB}/N_{L}^{XB}\lesssim 10^{-2}$ when $X\in\left\{T,E\right\}$ for \textit{all} experiments considered in Ref.~\cite{Grin:2011tf}, and so these errors are negligible on scales where the SU approximation is valid.

The curves shown there for $N_{L}^{XY}$ were actually numerically obtained using inverse-variance weighting of different multipole pairs, instead of the correct inverse-covariance weighting.  Inverse-variance weights are correct (and agree with inverse-covariance weights) for reconstruction noise from the pairs $TT$, $EE$, $TB$, and $EB$, for which the observable members of a pair are either totally correlated or uncorrelated in the absence of an isotropy-breaking realization of the CIP field. For $TE$, however, the neglect of covariance between $T$ and $E$ leads to incorrect behavior. Additionally, denominators in expressions for $\left(N_{L}^{TE}\right)^{-1}$ were evaluated as if $l=l'$ even when this was not the case. For $L\lesssim 100$, we find that this leads to errors of $\Delta N_{L}^{TE}/N_{L}^{TE}\lesssim 10^{-1}$ for \textit{all} experiments considered in Ref.~\cite{Grin:2011tf}, and so these errors are negligible on scales where the SU approximation is valid. 
\begin{figure*}[ht]
\includegraphics[scale=0.40]{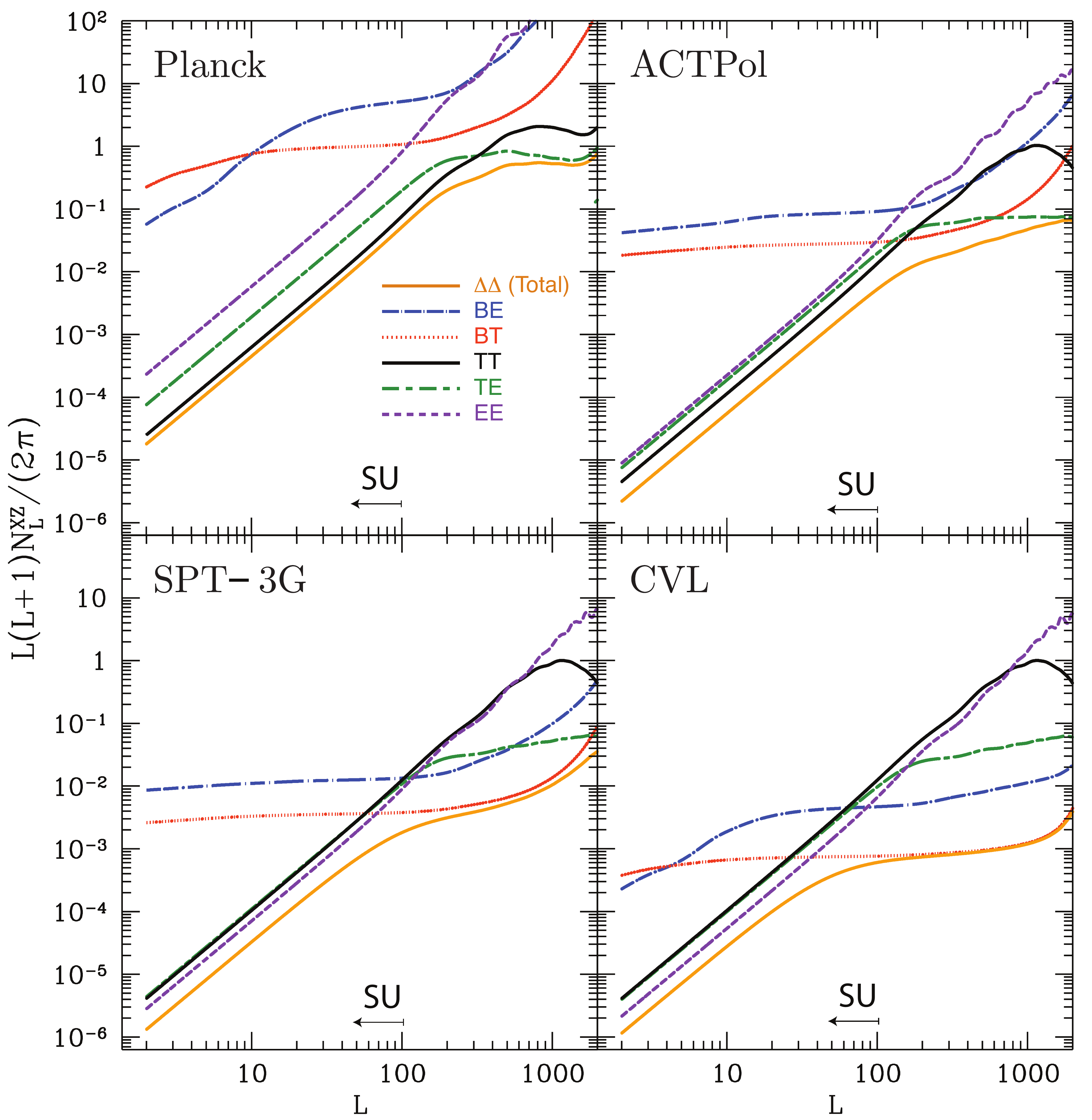}
\caption{Reconstruction noise curves $N_{L}^{XZ}$ for Planck, ACTPol, SPT-3G, and the cosmic-variance limit (CVL), obtained from the CIP estimator of Eq.~(\ref{eq:single_estimator}) with the indicated pair of observables $XZ$.  The approximate domain of validity of the SU approximation $L \lesssim 100$ is
indicated by the arrow. We assume the estimators employ 
CMB multipoles up to $l_{\rm CMB}=2500$.\label{fig:nc3}}
\end{figure*}

Calculations in Ref.~\cite{Grin:2011tf} were sped up by using permutation symmetries inside sums to simplify evaluations of $N_{L}^{XY}$. For $N_{L}^{TB}$, $N_{L}^{TE}$, and $N_{L}^{EB}$, summands were erroneously multiplied by a factor of $1/2$ when $l\geq |L-l'|$, but we find this leads to errors of $\Delta N_{L}^{TE}/N_{L}^{TE}\lesssim 2 \times 10^{-1}$ when $L<10$ and $\Delta N_{L}^{TE}/N_{L}^{TE}\lesssim 5 \times 10^{-2}$ when $10<L<1000$, and so these errors are small on scales where the SU approximation is valid.

For clarity and future reference, correct reconstruction noise curves $N_{L}^{XY}$ are shown in  {Fig. \ref{fig:nc3} for the Planck, ACTPol, and SPT-3G experiments, as well as for the CVL case. Reconstruction noise curves for CMB-S4 are shown earlier in this paper, in Fig. \ref{fig:nc}.} Total reconstruction noise curves are obtained using the full inverse-covariance weighting of different estimators [See Eqs.~(\ref{eq:full_estimator}) and (\ref{eq:covmat_total})] rather than the inverse-variance weighted sum used in Ref.~\cite{Grin:2011tf}.\footnote{In practice, this distinction only matters for $L$ ranges where the identity of the lowest-noise individual CIP estimator (e.g. $TT$, $EE$, $TE$, $TB$, or $EB$) is transitioning from one observable pair to another.} 

Using these reconstruction noise curves, we evaluate the signal-to-noise ratio for a detection of a scale-invariant spectrum of CIPs (with damping from projection as in Ref.~\cite{Grin:2011tf} for scales below the thickness of the recombination era) using only $\Delta \Delta$ auto-correlations, analogously to Ref.~\cite{Grin:2011tf}, but using Planck 2013 parameter values~\cite{Ade:2013uln}. Results are shown in Tables~\ref{tab:autotable_noproject} and~\ref{tab:autotable_project} for a variety of experiments and for the CVL case, with and without the SU domain of validity imposed, and with/without damping from projection. We now discuss the results qualitatively.
\begin{table*}
\caption{ Signal-to-noise ratio for a detection of a scale-invariant (angular) CIP power spectrum as a function of $\Delta_{\rm cl}^{2}$, the variance of the baryon fraction on galaxy-cluster scales. Shown below is the constant $\mathcal{C}$ in the relationship ${\rm SNR}=\mathcal{C}\times 1000\times\Delta_{\rm cl}^{2}$, where ${\rm SNR}$ is the signal-to-noise ratio for a CIP detection, with and without the domain of validity for the SU approximation imposed, with the reconstruction noise curve code of Ref.~\cite{Grin:2011tf} or the improved methods here. Here, damping due to projection for scales smaller than the thickness of the recombination era is included as in Ref.~\cite{Grin:2011tf}.}
\label{tab:autotable_noproject} 
\begin{ruledtabular}
\begin{tabular}{ccccc}
          Data &~~~~$\mathcal{C}$ (SU limit, improved $N_{L}^{\Delta \Delta}$)&~$\mathcal{C}$ (SU limit, old $N_{L}^{\Delta \Delta}$)&~$\mathcal{C}$ (no SU limit, improved $N_{L}^{\Delta \Delta}$)&~$\mathcal{C}$ (no SU limit, old $N_{L}^{\Delta \Delta}$)\\ \hline     	\noalign{\smallskip}

Planck   &~~~~8.73   &   8.58 & 8.73 &  8.58\\ 
ACTPol   &~~~~  13.8 & 13.6  &  13.8&13.9\\ 
SPT-3G   &~~~~19.2     & 19.1 &19.6&23.6\\ 
CMB-S4   &~~~~115      &   104  &117&128 \\ 
CVL      &~~~~171        & 155&193&226
\end{tabular}
\end{ruledtabular}
\end{table*}	

When erroneous reconstruction noise curves are used (with regard to the errors enumerated above), the spurious improvement obtained by neglecting the SU approximation can be as high as $\sim 50\%$ (in the CVL case). If, on the other hand, the improved reconstruction noise curves are used, the spurious improvement falls to $5\%$ for the CVL case. Put another way, if signal-to-noise ratio is evaluated including only modes for which the SU approximation is valid, the difference between the old and new noise curve codes is negligible. 

\begin{table*}
\caption{Same as Table~\ref{tab:autotable_noproject} but for no damping due to projection for scales smaller than the thickness of the recombination era. }
\label{tab:autotable_project} 
\begin{ruledtabular}
\begin{tabular}{ccccc}
          Data &~~~~$\mathcal{C}$ (SU limit, improved $N_{L}^{\Delta \Delta}$)&~$\mathcal{C}$ (SU limit, old $N_{L}^{\Delta \Delta}$)&~$\mathcal{C}$ (no SU limit, improved $N_{L}^{\Delta \Delta}$)&~$\mathcal{C}$ (no SU limit, old $N_{L}^{\Delta \Delta}$)\\ \hline     	\noalign{\smallskip}
 
Planck   &~~~~9.17   &   9.00 & 9.17 &  9.01\\ 
ACTPol   &~~~~  14.1 & 13.9  &  14.1&14.8\\ 
SPT-3G   &~~~~19.7     & 19.6 &20.1&32.7\\ 
CMB-S4   &~~~~120      &   110  &123&174 \\ 
CVL      &~~~~179        & 162&213&321
\end{tabular}
\end{ruledtabular}
\end{table*}	

If CIP projection damping is neglected (that is, a scale-invariant power spectrum of CIPs is assumed on all scales), the spurious improvement obtained by neglecting the SU approximation can be as high as $100\%$ (in the CVL) if the erroneous reconstruction noise curves are used. If, on the other hand, the improved reconstruction noise curves are used, the spurious improvement falls to $\sim 20\%$ for the CVL case. In all cases, the spurious improvement in signal-to-noise (for high $L_{\rm max}$) is caused by the polarization-driven flattening in the reconstruction noise curves at high $L$.

We recommend using the reconstruction noise curves and tables here for future CIP forecasts in the auto-only case. The differences between these different scenarios are much more dramatic if information from cross-correlations between $\Delta$ and $T$ or $E$ is included, as is the case in the rest of the paper.

\bibliography{chen_spires}
\end{document}